\documentclass[fleqn,usenatbib]{mnras}

\usepackage{mathptmx}

\usepackage[T1]{fontenc}

\DeclareRobustCommand{\VAN}[3]{#2}
\let\VANthebibliography\thebibliography
\def\thebibliography{\DeclareRobustCommand{\VAN}[3]{##3}\VANthebibliography}

\usepackage{graphicx}	
\usepackage{amsmath}	
\usepackage{amssymb}	
\def \LGalaxies{\texttt{L-Galaxies}\,}

\def\msun{\,\rm{M_\odot}}

\title [n-Hz GWB from galaxy formation models]{Massive black hole evolution models confronting the n-Hz amplitude of the stochastic gravitational wave background}

\author[Izquierdo-Villalba et al.]{
David Izquierdo-Villalba,$^{1,2}$\thanks{E-mail: david.izquierdovillalba@unimib.it}
Alberto Sesana,$^{1}$
Silvia Bonoli$^{3,4}$
and Monica Colpi$^{1,2}$
\\
$^{1}$Dipartimento di Fisica ``G. Occhialini'', Universit\`{a} degli Studi di Milano-Bicocca, Piazza della Scienza 3, I-20126 Milano, Italy\\
$^{2}$INFN, Sezione di Milano-Bicocca, Piazza della Scienza 3, 20126 Milano, Italy\\
$^{3}$ Donostia International Physics Centre (DIPC), Paseo Manuel de Lardizabal 4, 20018 Donostia-San Sebastian, Spain\\
$^{4}$IKERBASQUE, Basque Foundation for Science, E-48013, Bilbao, Spain\\
}

\date{Accepted XXX. Received YYY; in original form ZZZ}

\pubyear{2015}

\begin{document}
\label{firstpage}
\pagerange{\pageref{firstpage}--\pageref{lastpage}}
\maketitle

\begin{abstract}

We estimate the amplitude of the nano-Hz stochastic gravitational wave background (GWB) resulting from an unresolved  population of inspiralling massive black hole binaries (MBHBs).  To this aim, we use the \LGalaxies semi-analytical model applied on top of the \texttt{Millennium} merger trees. The dynamical evolution of MBHBs includes dynamical friction, stellar and gas binary hardening, and gravitational wave feedback. At the frequencies  proved by the \textit{Pulsar Timing Array} experiments, our model predicts an amplitude of ${\sim}1.2\,{\times}\,10^{-15}$ at ${\sim}\,3\,{\times}\,10^{-8}\, \rm Hz$ in agreement with current estimations. The contribution to the background comes primarily from equal mass binaries with chirp masses above $\rm 10^{8}\, M_{\odot}$. We then consider the recently detected common red noise in NANOGrav, PPTA and EPTA data,
working under the hypothesis that it is indeed a stochastic GWB coming from MBHBs. By boosting  the massive black hole growth via gas accretion, we show that our model can produce a signal with an amplitude $A\approx 2-3 {\times}\,10^{-15}$. There are, however, difficulties in predicting this background level without mismatching key observational constraints such as the quasar bolometric luminosity functions or the local black hole mass function. This highlights how current and forthcoming gravitational wave observations can, for the first time, confront galaxy and black hole evolution models.
\end{abstract}




\section{Introduction}


Due to recent major advances in the observational studies of Active Galactic Nuclei (AGN), evidence is growing that massive black holes (MBHs) heavier than  $10^5\msun $ form in nature and power AGN activity at the centers of  galaxies through gas accretion \citep{Schmidt1963,MerloniANDHeinz2008,Ueda2014,Hopkins2007,Aird2015}.  The demographic study of AGN and the dynamics of stars and gas around the center of nearby galaxies, further provided evidence that most (if not all) massive galaxies in the Universe host MBHs in their nuclei \citep{Genzel1987,Kormendy1988a,Dressler1988,Kormendy1992,Genzel1994,Salucci1999,Peterson2004,Vestergaard2006}. Even more, the existing correlations between the mass of MBHs and 
key properties of their host galaxies hint  for their co-evolution  \citep{Haehnelt1993,Faber1999,ODowd2002,HaringRix2004,Kormendy2013,Savorgnan2016}. 
Even though these findings sharpened our knowledge on the role of MBHs in the formation and evolution of galaxies, there is the need to contextualize galaxies and MBHs within the broad cosmological context. It is commonly accepted that the Universe behaves in a hierarchical way. Cosmic structure formed through the hierarchical assembly of dark matter (DM) halos, and the  galaxies observed nowadays assembled through mergers with smaller companions and accretion of matter from the cosmic filaments \citep{WhiteandRees1978,WhiteFrenk1991,Haehnelt1993,Kauffmann1999,Guo2011,Schaye2015,Vogelsberger2014a,Vogelsberger2014b,Nelson2018,Pillepich2018a}. Consequently, the existence of MBHs at the center of galaxies and the main role of mergers in the Universe, hint for the existence of  {\it massive black hole binary systems} (MBHBs) which might have formed and coalesced throughout the whole Universe lifetime.\\


Discovering the population of MBHBs is compelling but detecting dual or binary AGN  over a wide mass spectrum and redshift space is still a challenge  (see, for a review  \citealt{DeRosa2019}). An alternative avenue to discover MBHBs is provided by General Relativity. According to the theory, in fact,
MBHBs are  sources of gravitational waves (GWs), with frequencies ranging from above $10^{-9}\,\rm Hz$  up to  a few $10^{-2} \,\rm Hz$  \citep{Sathya2009,Colpi2017}. 
At the lowest frequencies around $10^{-9}\,{-}\,10^{-7} \, \rm Hz$, Pulsar-Timing Array experiments (PTA) aim at detecting the GW signal from a population of  MBHBs with masses around $10^{8-10}\msun$, thousands to millions of years prior to coalescence \citep{Sazhin1978,Foster1990,Rajagopal1995,JaffeBacker2003,Wyithe2003,Sesana2004,Enoki2004}. Although PTA experiments are also sensitive to GWs from single MBHBs, the most likely signal to be detected first is a stochastic gravitational background (GWB) produced by the incoherent superposition of GWs  from the cosmic population of inspiralling MBHBs out to $z\sim 1$ \citep{Rosado2015}. To detect such signal, PTA experiments search for spatial correlated fluctuations in the times of arrival of radio pulses from a network of millisecond pulsars in the Milky Way. 
Currently, three main PTA experiments are taking data: the \textit{European Pulsar Timing Array} \citep[EPTA,][]{Kramer2013,Desvignes2016}, the \textit{North American Nanohertz Observatory for Gravitational Waves} \citep[NANOGrav,][]{McLaughlin2013,Arzoumanian2015} and \textit{Parkes Pulsar Timing Array} \citep[PPTA,][]{Manchester2013,Reardon2016} projects. The three collaborations share data under the aegis of the International PTA \citep[IPTA,][]{Hobbs2010,Perera2019}. The final goal is to construct a global PTA with all the data collected around the world, including those provided by recently formed PTAs -- such as the Indian PTA \citep[InPTA,][]{Susobhanan2021} and the Chinese PTA \citep[CPTA,][]{Lee2016} --  and by cutting-edge new  timing instruments like MeerKAT \citep{Bailes2016}. In the last decade, EPTA, NANOGrav, PPTA and IPTA have been collecting data of ever improving quality, publishing a number of upper limits to the amplitude of the GW background $A_{\rm yr^{-1}}\,{\lesssim}\,2-3\,{\times}\,10^{-15}$ at $\rm 1\, yr^{-1}$ \citep{Lentati2015,Shannon2015,Verbiest2016,Arzoumanian2018}.
Interestingly, the most recent results of NANOGrav (12.5-year data set), PPTA (second data release, DR2) and EPTA (DR2) have pointed out the existence of a stochastic process with median amplitude of $A_{\rm yr^{-1}}\,{\sim}\,(1.9-2.95)\,{\times}\,10^{-15}$ \citep{Arzoumanian2020,Goncharov2021,Chen2021}. However, the lack of significant evidence of the quadrupolar correlations in such detected signals makes difficult to claim a GWB detection.\\


From a theoretical point of view, several works  aim at predicting the expected stochastic GWB at nHZ frequencies. For instance, \cite{Jaffe2003} reported an amplitude of $A_{\rm yr^{-1}}\,{\sim}\,10^{-16}$ by linking the observed merger rate of massive galaxies with some analytical prescriptions for MBH binary evolution. However, \cite{Wyithe2003} showed that $A_{\rm yr^{-1}}$ could increase up to ${\sim}\,10^{-15}$ if the galaxy merger rate is computed from the extended Press \& Schechter theory \citep[PS,][]{PressANDSchechter1974}. Such discrepancies were principally due to the different analytical recipes used to treat the DM halo and black hole physics, which in turn reflected the lack of knowledge about how halos and MBH binaries co-evolve with cosmic time. Indeed, the large variance caused by such effect was noticed by \cite{Sesana2008} who carried out a systematic study on the GW stochastic background predicted by a wide variety of semi-analytical models (SAMs) based on the PS halo mass function.   
The authors concluded that taking into account the uncertainties of all these models, the expected GWB amplitude detected by PTA could expand between $A_{\rm yr^{-1}}\,{\sim}\, 2.4\,{\times}\,10^{-16}$ and $A_{\rm yr^{-1}}\,{\sim}\,3.8\,{\times}\,10^{-15}$. To improve the statistics of the PS halos and to avoid the overproduction of low-$z$ bright quasar seen in PS-based models \citep[e.g.][]{Marulli2006} a number of works used merger trees extracted form cosmological N-body simulation. Among them, we cite \cite{Sesana2009}, which explored the PTA predictions using the catalogue of merging galaxies extracted from \cite{Bertone2007} semi-analytical model applied on the \texttt{Millennium} DM merger trees \citep{Springel2005}. By associating to each merging galaxy a central MBH according to some observational prescription, the authors reported $4{\times}\,10^{-16}\,{<}\,A_{\rm yr^{-1}}\,{<}\, 2{\times}\,10^{-15}$. Besides, \cite{Sesana2009} concluded that depending on the model used for placing MBHs, individual signals from MBHBs could be detected in PTA data. However, these types of events are likely to be rare. Similar work was performed by \cite{Roebber2016} using the N-body simulations \texttt{Dark Sky} and \texttt{MultiDark} \citep{Riebe2011,Skillman2014}: placing galaxies and MBHs inside DM halos through scaling relations and leaving aside a detailed modelling of binary dynamics and associated delay after the halo-halo merge, the authors found a typical value of $A_{\rm yr^{-1}}\,{\sim}\,6{\times}\,10^{-16}$. Another class of models directly exploits observations of galaxy pairs to infer a galaxy and MBHB merger rate, which is then used to construct a stochastic GWBs. Such models were extensively investigated by \cite{Sesana2013,2015MNRAS.447.2772R} and \cite{Sesana2016}, yielding  $3\,{\times}\,10^{-16}\,{<}\,A_{\rm yr^{-1}}\,{<}\, 2\,{\times}\,10^{-15}$, due to uncertainties in defining galaxy pairs, estimating merger timescales and connecting MBHs to their hosts via a bulge-MBH mass relations \citep{Kormendy2013,Shankar2016}.

Even though all these models were already providing strong constraints on the GWB at nHz frequencies, they relied on uncertain observations and/or empirical relations to place galaxies and MBHs in the DM merger trees. Crucially, they missed a self consistent treatment of galaxy evolution and of how MBHs and MBHB form and evolve inside galaxies. To improve these limitations, \cite{Dvorkin2017} and \cite{Bonetti2018ModelTriplets} based their GWB predictions on the SAM of \cite{Barausse2012}. This model, based on Press \& Schechter merger trees, had the advantage of including a detailed modelling for the cosmological evolution of galaxies and MBHs, and it further refined to include different prescriptions for the MBH binary evolution. On one side, \cite{Dvorkin2017} explored the GWB amplitude within the PTA band in the worst possible scenario, i.e., if all MBHBs are not able to merge and they are stalled at ${\sim}\, \rm pc$ scales (i.e the so called \textit{final-parsec problem}, \citealt{Milosavljevic2001}).  Their results showed that even in this pessimistic scenario, a GW signal should remain in the PTA band ($A_{\rm yr^{-1}}\,{\sim}\,10^{-16}$). On the other hand, \cite{Bonetti2018ModelTriplets} performed a similar study but extending the treatment of MBH binaries and including a refined model of triple MBH interactions as a plausible mechanism for avoiding the stalling of MBHBs. The authors reported $A_{\rm yr^{-1}}\,{\sim}\,10^{-15}$ and highlighted that  triple interactions between a MBHB and a  MBH orbiting around the binary or impinging on it,  play an important role in the final GWB amplitude, avoiding the reduction of the signal as a consequence of the stalling binaries. Thanks to the fast development of cosmological hydrodynamical simulations able to follow the assembly of galaxies down to relatively small scales in large cosmological volumes, recent works have also drawn predictions for $A_{\rm yr^{-1}}$ by taking advantage of the galaxy properties provided by these simulations. \cite{Kelley2016} used the galaxy population of the \texttt{Illustris} simulation \citep{Vogelsberger2014a,Vogelsberger2014b} to construct a comprehensive modelling for tracking the different evolutionary stages of MBH binaries. With such \textbf{a} model, \cite{Kelley2016} reported $A_{\rm yr^{-1}}\,{\sim}\,7\,{\times}\,10^{-16}$ with most of the signal coming from very massive binaries (${\sim}\, \rm 10^9 \, M_{\odot}$) merging at low-$z$ ($z\,{<}\,3$). This work was extended by \cite{Siwek2020}, which explored the repercussion of gas accretion in MBH binaries on the GWB. Their results showed that if the growth of the secondary MBH is favored, the GWB level could reach up to $A_{\rm yr^{-1}}\,{\sim}\,10^{-15}$. On the contrary, in the case in which the secondary MBH growth is halted, the GWB dropped down to $A_{\rm yr^{-1}}\,{\sim}\,3\,{\times}\,10^{-16}$.\\

A fact worth noticing is that  GWB amplitudes up to $A_{\rm yr^{-1}}\,{\approx}\,4\times 10^{-15}$ can be found in the literature. However, models that self consistently evolve galaxies and MBHs and that  reproduce the MBH mass and quasar luminosity functions hardly get a GWB level much in excess of  $A_{\rm yr^{-1}}\,{\approx}\,1\times 10^{-15}$ \citep{Kelley2017,Bonetti2018ModelTriplets}. This is particularly interesting in light of the recent results of the NANOGrav (12.5-year data set), PPTA (DR2) and EPTA (DR2) collaboration which reported strong evidences of a stochastic process with $A_{\rm yr^{-1}}$ spanning respectively, between $1.37\,{\times}\,10^{-15} \,{-}\, 2.67\,{\times}\,10^{-15}$, $1.9\,{\times}\,10^{-15} \,{-}\, 2.6\,{\times}\,10^{-15}$ and $2.23\,{\times}\,10^{-15} \,{-}\, 3.8\,{\times}\,10^{-15}$ \citep{Arzoumanian2020,Goncharov2021,Chen2021}. Note that the signal seen by NANOGrav, PPTA and EPTA did not display significant evidence of the quadrupolar correlations needed to claim  detection of a GWB. Nevertheless, 
it is fundamental to explore what theoretical models can tell us about such a large signal level. 
Even more, precision in the measurement of the GWB amplitude could be used as a new tool to improve our knowledge about the co-evolution of MBHs and galaxies, rule out theoretical models of MBH binary evolution and test our current treatment of galaxy formation, whose detailed modelling is still a challenge. \\

Motivated by this, in this work we explore the evolution of MBH binaries in the context in galaxy formation models. For that we introduce a model of MBHB formation and evolution embedded inside the \LGalaxies semi-analytical model in the version of \cite{IzquierdoVillalba2019,IzquierdoVillalba2020}.
Specifically, unlike many other SAMs in the literature, the model introduces recipes for the MBH dynamics in the host galaxy, as the MBHB coalescence is not instantaneous. The MBHs need to reach subparsec scales for gravitational waves to drive the evolution and enter the PTA bandwidth. Thus, stellar and gas dynamical torques acting  on galactic scales lead to delays in the computation of the binary merger timescale compared to the timescale of the colliding galaxies.
All these processes have been included in a self-consistent manner inside the cosmological evolution of galaxies and black holes tracked by \LGalaxies. We have applied the new model on the \texttt{Millennium} DM merger trees \citep{Springel2005} whose box-size and mass resolution had offered us the capability of drawing predictions for GW emission in the PTA band. 
To our knowledge, this work is the first to include current GWB measurements as an extra constraint to calibrate the evolution of MBHs within the context of galaxy formation models. In particular, we add the GWB to the standard constraints provided by the quasars luminosity function (QLF) and mass function of MBHs (BHMF) in the local Universe \citep{Marconi2004,Hopkins2007, Shankar2009,Shen2020}. In this way, we are able to explore, for the first time, how feasible is for these galaxy formation models (and in particular, our version of \LGalaxies) to jointly reach the current measurements of the GWB while reproducing the well constrained QLF and BHMF.

The paper is organised as follows: In Section~\ref{sec:SAM_MILL} we describe the main characteristics of \LGalaxies and \texttt{Millennium} simulations. In Section~\ref{sec:MBHBModel} we present the model that traces the formation and evolution of MBHBs. In Section~\ref{sec:Results} we present our results, focusing on the GW signal in the PTA frequency band and the difficulties of the model to produce large GW amplitudes without mismatching other MBH constrains such as MBH mass function. A Lambda Cold Dark Matter $(\Lambda$CDM) cosmology with parameters $\Omega_{\rm m} \,{=}\,0.315$, $\Omega_{\rm \Lambda}\,{=}\,0.685$, $\Omega_{\rm b}\,{=}\,0.045$, $\sigma_{8}\,{=}\,0.9$ and $\rm H_0\,{=}\,67.3\, \rm km\,s^{-1}\,Mpc^{-1}$ is adopted throughout the paper \citep{PlanckCollaboration2014}.\\


\section{Galaxy formation model} \label{sec:SAM_MILL}
In the following sections, we   briefly overview the main physics included in the \LGalaxies semi-analytical model.  \LGalaxies $\,$ is a code that tracks the time evolution of gas, stars and MBHs  within their host dark matter subhalos\footnote{In this work we  define subhalos as locally overdense, self-bound particle groups formed inside the DM halos.} through a series of differential equations and analytic prescriptions. The version of the model used here is the \cite{Henriques2015} but with the modifications in the bulge and black hole physics presented in \cite{IzquierdoVillalba2019,IzquierdoVillalba2020}. 

\subsection{Dark matter merger trees}
DM merger trees are the backbone of any semi-analytical model. In this paper we  use the trees extracted from the \texttt{Millennium} N-body simulation \citep[hereafter MS,][]{Springel2005}. MS follows the cosmological evolution of $2160^3$ DM particles with a mass of $8.6 \times 10^8\, \mathrm {M_{\odot}}/h$ within a periodic cube of 500 ${\rm Mpc}/h$ on a side. Even though MS was run by using WMAP1 \& 2dFGRS cosmology, the version of \LGalaxies in this work is tuned on a re-scaled versions of the MS simulation \citep{AnguloandWhite2010} to match the cosmological parameters obtained by Planck first-year data release  \citep{PlanckCollaboration2014}.\\

All the particle information of MS is stored at 63 different epochs or \textit{snapshots}. 
At every snapshot DM halos and subhalos are extracted  using a friend-of-friend (FOF) group-finder and \texttt{SUBFIND} algorithm \citep{Springel2001}, respectively. By applying \texttt{L-HALOTREE} \citep{Springel2005} all halo and subhalo structures are  arranged in merger trees to follow the evolutionary path of any DM (sub)halo in the simulations. We highlight that \LGalaxies is based on the subhalo population instead of the halo one. This enables \LGalaxies $\,$ to build-up a more realistic galaxy population, making more reasonable predictions on the galaxy merger rate and clustering. Nevertheless, the time resolution given by the 63 snapshots is not enough to properly trace the baryonic physics. Thus, the SAM does an internal time discretization between two consecutive snapshots with approximately $\rm{\sim}\,5{-}20 \,Myr$ of time resolution. These extra-temporal subdivisions of \LGalaxies $\,$ are called \textit{sub-steps}. 

\subsection{Baryonic physics}\label{sec:barPhys}
\LGalaxies{} follows the standard scenario of structure formation, by assuming 
that when a subhalo virializes, part of the diffuse baryonic gas present  in its surroundings  is trapped and collapses within it. Baryons are deposited in the subhalo in the form of a hot gas atmosphere. Within the cooling timescale, this gas gradually migrates towards the center of the subhalo, forming a disk-like structure, called cold-gas disk. When the disk is large enough, episodes of star formation are triggered, leading to the assembly of the stellar disk. \LGalaxies  self-regulates the formation of stars by including feedback both from a central AGN and supernovae.  Galaxies are able to form a over-density of stars in the nuclear region (i.e the so-called bulge) via mergers and disk instabilities (DI). According to the baryonic merger ratio of the two interacting galaxies, the remnant can be transformed into an elliptical galaxy, or can preserve the stellar disk developing a galactic bulge by incorporating the whole stellar component of the smaller  progenitor. In the model used here, we introduce the concept of \textit{smooth accretion}, which occurs when the less massive progenitor is completely absorbed by the stellar disk of the central galaxy \citep{IzquierdoVillalba2019}. Alternatively, disk instabilities
in massive disks can change the stellar distribution, leading to the formation of a central ellipsoidal component, typically referred to as bar or pseudo-bulge.

\subsection{Black hole physics: growth and spin} \label{sec:BlackHoleGrowthSpin}

Each newly resolved subhalo (independently of redshift and halo properties) is seeded with a black hole of $\rm 10^{4} M_{\odot}$ whose spin has a modulus $\vert a\vert $ randomly selected between $0\,{<}\,|a|\,{<}\,0.998$. The choice of the initial seed mass is  conservative  given the minimum mass of new resolved subhalos in the MS (${\sim}\,10^{10}\rm M_{\odot}$). In future works we will explore the model predictions for MBHs  using the refined seeding procedure presented in Spinoso et al. (in preparation).  Once the black hole seed is placed in its host galaxy, it can grow through three different channels: \textit{cold gas accretion}, \textit{hot gas accretion} and \textit{mergers} with other black holes. Specifically, the first channel is the main driver of the black hole growth and it is triggered by both galaxy mergers  and disk instability events. After a galaxy merger we assume that the fraction of cold gas accreted by the nuclear black hole is:
\begin{equation}\label{eq:QuasarMode_Merger}
\rm   \Delta {M}_{BH}^{gas} \,{=}\,\mathit{f}_{BH}^{merger} (1+\mathit{z}_{merger})^{5/2} \frac{m_{R}}{1 + (V_{BH}/V_{200})^2}\, M_{\rm gas},
\end{equation}
where $\rm m_{R}\,{=}\,M_{satellite}^{baryon}/M_{central}^{baryon}\,{\leq}\,1$ is the baryonic ratio of the two interacting galaxies, $\rm V_{200}$ the virial velocity of the host DM subhalo, $z_{\rm merger}$ the redshift of the galaxy merger, $\rm M_{\rm gas}$ the cold gas mass of the galaxy and $\rm V_{BH}$, $\rm \mathit{f}_{BH}^{\rm merger}$ two adjustable parameters set to $\rm 280 \, km/s$ and $0.025$, respectively. In presence of a disk instability the black hole accretes an amount of cold gas proportional to the mass of stars that trigger the stellar disk instability, $\rm \Delta M_{\rm stars}^{DI}$\footnote{Disk instabilities are accounted for by \LGalaxies using the \cite{Efstathio1982} criterion. Based on that prescription, the amount of matter which triggers a disk instability event is set to:
\begin{equation*}\label{eq:Delta_M_stars_DI}
 \rm \Delta M_{stars}^{DI} \,{=}\, M_{\rm \star, d} - \left( V_{max}^2 R_{\star,d}/G \epsilon^{2}\right)  \, {>}\,0 \, ,
\end{equation*}
where $\epsilon$ is a free parameter set to $1.5$, $\rm V_{max}$ is the maximum circular velocity of the host dark matter, $\rm R_{\star,d}$ and  $\rm M_{\star,d}$ are  the length and stellar mass of the stellar disk, respectively.}:
\begin{equation}\label{eq:QuasarMode_DI}
\rm    \Delta {M}_{BH}^{gas} \,{=}\, \mathit{f}_{BH}^{DI} (1+\mathit{z}_{DI})^{5/2} \frac{\Delta M_{stars}^{DI}}{{1 + (V_{BH}/V_{200})^2}},
\end{equation}
where $\rm \mathit{z}_{DI}$ is the redshift in which the disk instability takes place, and $\rm \mathit{f}_{BH}^{DI}$ is a free parameter that takes into account the gas accretion efficiency, set to $0.0015$.
We highlight that the redshift dependence of Eq.~\ref{eq:QuasarMode_Merger} and Eq.~\ref{eq:QuasarMode_DI} has been modified with respect to \cite{IzquierdoVillalba2020} to improve the match between the observed and the  predicted black hole mass function and bulge-MBH correlations at $z\,{=}\,0$ .\\

After a galaxy merger or a disk instability, the cold gas available for accretion is assumed to settle in a reservoir around the black hole, $\rm M_{Res}$. Instead of an instantaneous gas consumption, the model considers that the gas reservoir is progressively consumed trough a Eddington-limited growth phase, followed by a second phase of low accretion rates \citep{Hopkins2005,Hopkins2006a,Marulli2006,Bonoli2009}. We refer the reader to \cite{IzquierdoVillalba2020} for further details.\\ 

During any of the events that make the MBH grow, the code tracks the evolution of the black hole spin in a self-consistent way. During gas accretion events, the model uses the approach presented in \cite{Dotti2013} and \cite{Sesana2014}, which links the number of accretion events that spin-up or spin-down the MBH with the degree of coherent motion in the bulge. In particular, the model assumes that disk instabilities  increase the coherence of the bulge kinematics. On the other hand, mergers bring  disorder to the bulge dynamics. After a MBH coalescence the final spin is determined by the expression of \cite{BarausseANDRezzolla2009}, where a distinction between wet and dry mergers is done to compute the alignment/anti-alignment between the two MBHs. For further details on the implementation in the SAM, we refer the reader to \cite{IzquierdoVillalba2020}. We highlight that in this work we do not include the gravitational recoils after coalescence, as presented in \cite{IzquierdoVillalba2020}. In a future work we will explore what is the effect of recoils on the population of MBHBs.\\

Finally, we highlight that all the parameters used in the SAM (including the ones of Eq.~\ref{eq:QuasarMode_Merger} and Eq.~\ref{eq:QuasarMode_DI}) have been chosen to reproduce many observed galaxy and MBH properties. Among them, we can highlight the stellar mass function, the fraction of passive galaxies, quasar luminosity function, the $z\,{=}\,0$ black hole mass function or the $z\,{=}\,0$ correlation between bulge and black hole mass (we refer to \citealt{Henriques2015} and \citealt{IzquierdoVillalba2020} for the specific comparisons).

\section{The population of massive binary black holes} \label{sec:MBHBModel}

In this section we describe the physics included in \LGalaxies to follow the formation and coalescence of MBHBs. Following \cite{Begelman1980} we divide the evolutionary pathway of MBHBs into three stages. The first one is described in Section~\ref{sec:PairingPhase} and consists of a \textit{pairing} phase in which, after the galaxy merger, the dynamical friction exerted by the stars drives the MBH of the satellite galaxy toward the nucleus of the remnant galaxy where it binds with the central MBH. This occurs when the amount of stars enclosed within the binary orbit is comparable to the mass of the lighter MBH of the binary. Then, a \textit{hardening} phase takes place in which the orbital semi-major axis of the binary shrinks due to three-body interactions with single stars (the slingshots mechanism)  and/or interaction with a massive gaseous circumbinary disk \citep{Colpi2014}. Finally, a \textit{gravitational wave inspiral} phase drives the binary to coalescence. We discuss the implementation of the two last phases in Section~\ref{sec:HardeningModel}.\\ 

\subsection{The pairing phase of massive black holes} \label{sec:PairingPhase}

The first phase which anticipates the formation of a binary system at the center of the   post-merger galaxy consists in reducing the MBH separation from  ${\sim}\, \rm kpc$  to ${\sim}\, \rm pc$ through dynamical friction. In this work, to estimate the time spent by a black hole in the pairing phase, we use the expression \citep{BinneyTremaine2008}:
\begin{equation}  \label{eq:DynamicalFriction}
    t_{\rm dyn}^{\rm BH} \,{=} \, 19 \, f(\varepsilon)  \left( \frac{r_0}{5 \, \rm kpc} \right)^2 \left( \frac{\sigma}{200 \rm km/s}\right) \left( \frac{10^8 \, \rm M_{ \odot}}{\rm M_{BH}} \right) \, \frac{1}{\Lambda}\, \rm [Gyr] , 
\end{equation}
where $f(\varepsilon)$ is a function with depends on the orbital circularity of the black hole $\varepsilon$ \citep{Colpi1999}, $r_0$ is the initial position of the black hole deposited by the satellite galaxy after the merger, $\sigma$ is the velocity dispersion of the remnant galaxy ($\sigma^2\,{=}\, \rm 0.25 G M_{stellar}/R_{gal}$)\footnote{$\rm R_{gal}$ refers to the effective radius of the galaxy, computed as the mass weighted average of the galaxy bulge and stellar disc radius. In \cite{IzquierdoVillalba2019} it was showed that $\rm R_{gal}$ values predicted by \LGalaxies are compatible with current observations.}, $\rm M_{BH}$ is the mass of the black hole and $\rm \Lambda\,{=}\,\ln(1 + M_{stellar}/M_{BH})$ is the Coulomb logarithm  \citep{MoWhite2010}.\\

\begin{figure}
\centering
\includegraphics[width=1.\columnwidth]{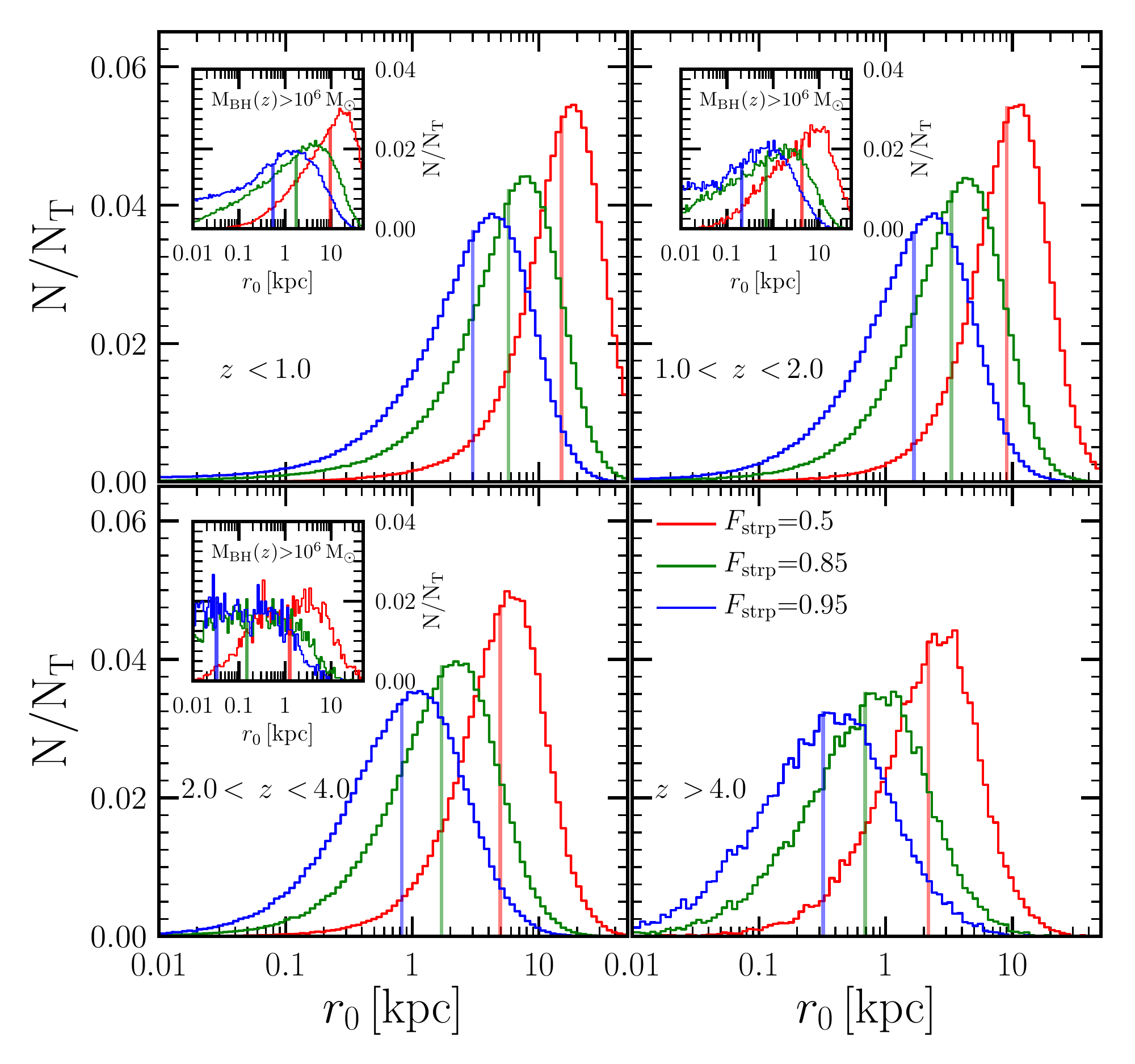}
\caption[]{Distribution of $r_0$ for three different values of $F_{\rm strp}$ representing the mass lost by the secondary due to tidal stripping by the primary galaxy: 0.5 (red), 0.85 (green) and 0.95 (blue). Solid vertical lines represent the median value for each distribution. Each panel displays a different redshift bin. The inner panels show the same but only for satellite galaxies which deposit a MBH of mass ${>}\,10^{6}\, \rm M_{\odot}$. We do not show the cases at $z\,{>}\,4$ given the small number of satellite galaxies with ${>}\,10^{6}\, \rm M_{\odot}$ MBHs at that high $z$.}
\label{fig:r0_BHs}
\end{figure}

The value of $r_0$ in Eq.~\ref{eq:DynamicalFriction} is the position where the satellite galaxy has lost a fraction $F_{\rm strp}$ of its total mass by tidal stripping. Such position is determined by solving numerically the expression \citep{King1962,TaylorBabul2001}:
\begin{equation}
    \frac{d^2\Phi(r)}{dr^2} \,{=} \, \omega^2 - \frac{\mathrm{G}\, \mathrm{M_{sat}}(<R)}{R^3},
\end{equation}
where the variable $r$ is the radial position of the satellite galaxy within the dark matter subhalo, $\omega$ is its instantaneous orbital angular velocity, and $\Phi$ the potential of the hosting subhalo. Finally $R$ and $\rm M_{sat}(<R)$ represent the radius and mass at which the satellite galaxy contains $(1\,{-}\,F_{\rm strp})$ of its total baryonic mass. While the value of $\rm M_{sat}(<R)$ is computed assuming exponential disk and Sérsic bulge profiles  \citep{Sersic1968}, the subhalo potential is modeled as a \textit{Navarro-Frenk-White} \citep[NFW,][]{NFW1996}\footnote{Given that the \texttt{Millennium} merger trees catalogues do not contain the subhalo concentration, we use the fits of \cite{Dutton2014} to obtain their concentration at any redshift and mass.}. Given the limitations of \LGalaxies to provide accurate positions of satellite galaxies which had lost their dark matter subhalo, we evaluate the quantities of Eq.~\ref{eq:DynamicalFriction} at the instant at which the DM subhalo associated with the satellite galaxy merges with the one associated with the central galaxy. From this moment, the DM host of the satellite is not resolved anymore by the DM simulation.\\

In Fig.~\ref{fig:r0_BHs} we present the distribution of $r_0$ for three different values of  $F_{\rm strp}$ (0.5, 0.85 and 0.95). As we can see, the larger is $F_{\rm strp}$ the smaller is $r_0$. Moreover, regardless of $F_{\rm strp}$, there is a redshift evolution in the $r_0$ values. In particular, the smaller is the redshift, the larger is the typical $r_0$. This is a consequence of the increase of DM halo mass and its concentration towards low redshifts \citep{Dutton2014}, which causes the halo potential to be more efficient in disrupting the satellite galaxy. To check if the $r_0$ distribution changes for the most massive MBHs, in the inner plots of Fig.~\ref{fig:r0_BHs} we present the values of $r_0$ only for satellite galaxies which deposit a $\rm {>}\,10^6 \, M_{\odot}$ MBH. 
As shown, these galaxies follow the general trend of the large $F_{\rm strp}$ values being associated with small $r_0$ values. Nevertheless, regardless of $F_{\rm strp}$, they have a median $r_0$ smaller than the general population. This deviation is caused because the former population have stellar masses ${\sim}\,1\, \rm dex$ larger: $\rm M_{stellar}\,{\sim}\,10^{9.5} \, M_{\odot}$ versus $\rm M_{stellar}\,{\sim}\,10^{8.7} \, M_{\odot}$ of the general satellite population. This mass difference causes that satellite galaxies hosting $\rm {>}\,10^6 \, M_{\odot}$ MBHs take more time before being stripped, having more chances to deposit the MBH at low $r_0$ values.  
In this work we decided to use $F_{\rm strp}\,{=}\,0.85$. Even though this choice is somewhat arbitrary, we selected such a high threshold to be sure that most of the stellar component around the satellite MBH is already tidally removed by the merging process. Thus, the dynamics of the MBH can be progressively considered as the one of a \textit{naked MBH} moving in the stellar background of the remnant galaxy.\\ 




\begin{figure}
\centering
\includegraphics[width=1.\columnwidth]{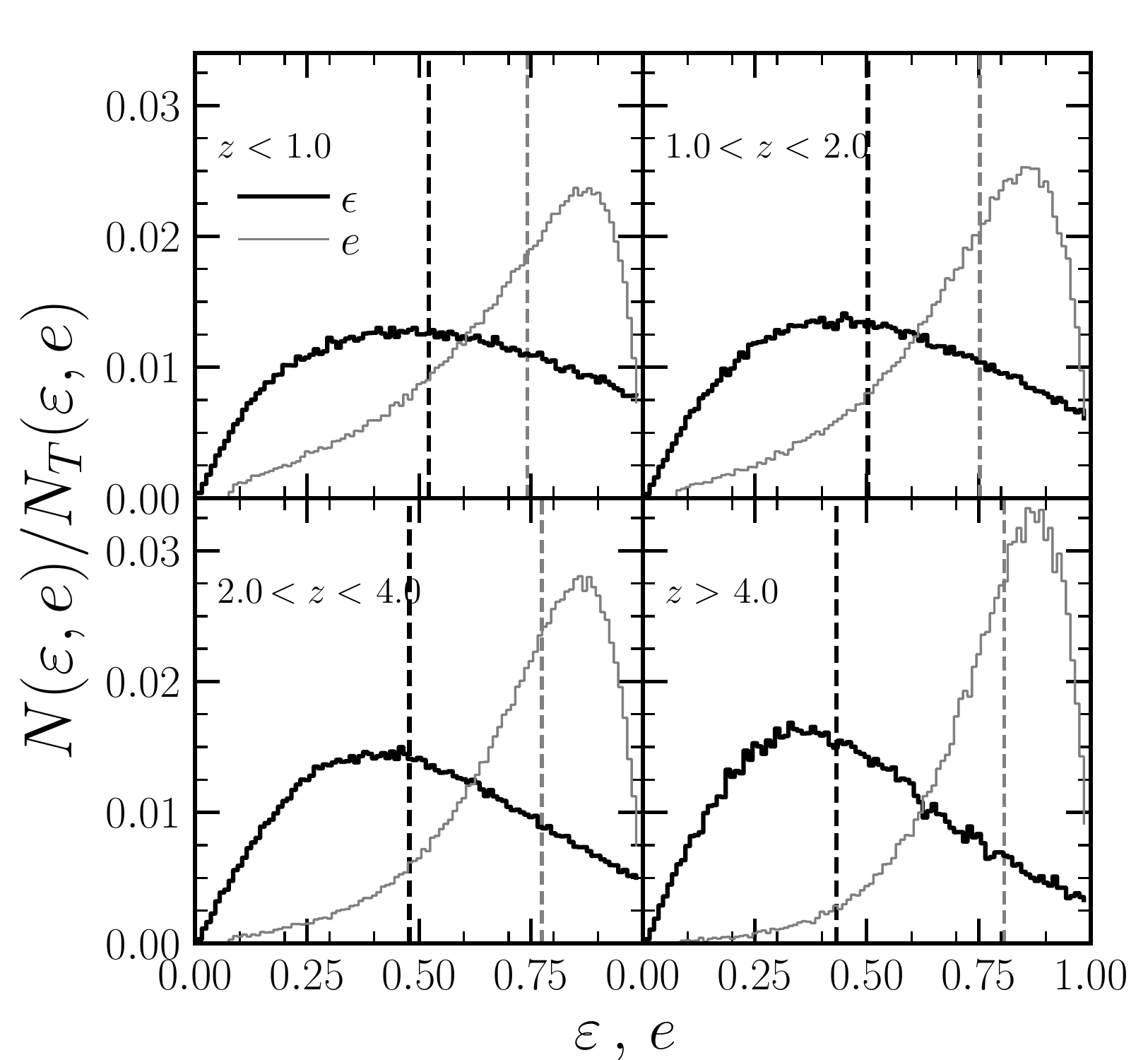}
\caption[]{Distribution of circularity ($\varepsilon$, thick black line) and eccentricity ($e$, thin grey line) of the satellite black holes at the moment in which they are deposited at $r_0$. Different panels represent different redshifts bins and vertical lines represent the median of the distribution.} 
\label{fig:Ecc_Circularity_t_dyn_ecc}
\end{figure}

As shown in Eq.~\ref{eq:DynamicalFriction}, the dynamical friction time scale depends on the circularity of the MBH orbit. Following \cite{Lacey1993} we adopt  $f(\varepsilon) \sim \varepsilon^{0.78}$ (see other methods as \citealt{Colpi1999,BoylanKolchin2008}). Here we assume that that the MBH orbital circularity is inherited form the one of the satellite galaxy. Specifically, galaxy orbital circularities are computed following \cite{Scannapieco2009} \citep[see also][]{Abadi2003} defining $\varepsilon$ as:
\begin{equation}\label{eq:circularity}
\varepsilon \,{=}\, \frac{j}{j_c},
\end{equation}
where $j$ is the angular momentum per unit of mass of the satellite galaxy at a distance $r$ from the halo center and $j_c$ the angular momentum expected for a circular orbit at the same $r$, i.e $j_c\,{=}\, r\,v_c(r) = r \sqrt{G \mathrm{M}_h(<r)/r}$, where $\mathrm{M}_h(<r)$ is the halo mass enclosed within $r$, computed assuming a NFW profile. As we did before, Eq.~\ref{eq:circularity} is computed as soon as the satellite galaxy looses its host DM subhalo. In Fig.~\ref{fig:Ecc_Circularity_t_dyn_ecc} we show the orbital circularity of the MBHs in the \texttt{Millennium} dark matter merger trees. As we can see, $\varepsilon$ has a moderate evolution with redshift. The peak around $\varepsilon\approx 0.35$ gets progressively smeared out, with larger circularities becoming more common at lower redshift.
We also show for completeness the distribution of orbital eccentricities\footnote{This value has been computed as $e\,{=}\,(r_+-r_-)/(r_+ + r_-)$, being $r_+$ and $r_-$ the apo- and peri- center of the orbit. Such quantities are the roots of $\left(1/r^2\right) + \left(2\left[\Phi(r)-E \right]/L^2\right) \,{=}\,0$. The values of $E$ and $L$ are, respectively, the energy and angular momentum per unit mass in a spherical potential ($\Phi$, in our case the NFW potential).}. As we move to lower redshifts, the distribution tends to develop a substantial tail at low values, although it maintains a maximum at $e\approx 0.8$, in agreement with \cite{Tormen1997}.\\

\begin{figure}
\centering
\includegraphics[width=1.\columnwidth]{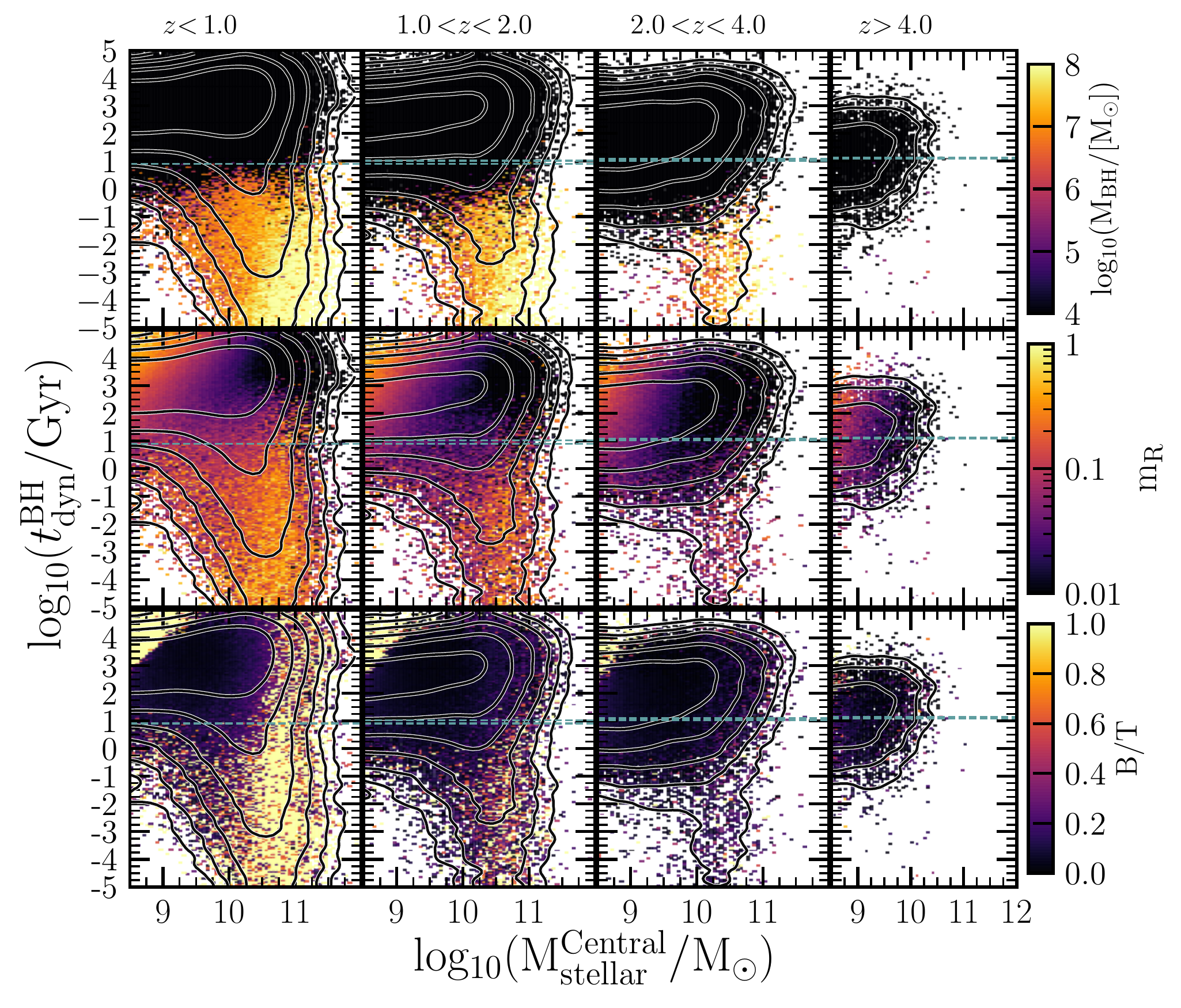}
\caption[]{$\rm M_{stellar^{\rm Central}}\,{-}\,t_{\rm dyn}^{\rm BH}$ plane at four different redshift bins ($z\,{<}\,1$, $1\,{<}\,z\,{<}\,2$, $2\,{<}\,z\,{<}\,4$ and $z\,{>}\,4$). The first row encodes the median black hole mass of the MBH in the pairing phase ($\rm M_{BH}$) at a given bin of $\rm M_{stellar}^{Central}$ and $\rm t_{\rm dyn}$. In the second row the color represents the baryonic galaxy merger ratio, $\rm m_R$. The color map of the third panel encodes the bulge-to-total ratio (B/T) of the hosting galaxy. In all the panels, the blue lines represent the Hubble time ($\rm t_{\rm H}$) at a given redshift bin. In all the panels, the white lines represent the contours where a same number of MBHs is enclosed it.}
\label{fig:t_dyn_BH_tidal_All_MS}
\end{figure}

Fig.~\ref{fig:t_dyn_BH_tidal_All_MS} carries information on key quantities in the plane $\mathrm{M_{stellar}^{Central}}\,{-}\,t_{\rm dyn}^{\rm BH}$, where the former is the     stellar content of the post-merger galaxy. We show the results for $\rm M_{stellar}^{Central}\,{>}\,10^{8.5} M_{\odot}$ corresponding to the range above which the results are not significantly affected by resolution of the underlying \texttt{Millennium} DM simulation. In each panel, the distribution has been color coded by the mass of the satellite MBH in the pairing phase ($\rm M_{BH}$), the baryonic merger ratio of the two interacting galaxies ($\rm m_R$) and the \textit{bulge-to-total} ratio of the remnant galaxy ($\rm B/T$). At $z\,{>}\,4$ there is a significant fraction of satellite MBHs (${\sim}\,47\%$) that would reach the center of the galaxy within the Hubble time ($t_{\rm H}$). This is principally caused by the fact that at high-$z$, DM subhalos are smaller and galaxies more compact. Due to the combination of these two facts, satellite galaxies are less affected by strong tidal effects and thus capable deposit the MBH at closer distances from the nucleus of the central galaxy (see $r_0$ of Fig.~\ref{fig:r0_BHs}). Interestingly, most of these MBHs are close to the seed mass ($10^4\, \rm M_{\odot}$), which is a direct consequence of the rough seeding procedure used in this work. On the other hand, at lower redshifts the situation changes and the number of MBHs with $t_{\rm dyn}^{\rm BH}\,{>}\,t_{\rm H}$ is the predominant (${>}\,80\%$ of the cases). Even though the fraction of MBHs which merge in a Hubble time is decreasing, essentially all systems with a MBH larger than a $\rm 10^6 \, M_{\odot}$ do so. Indeed, all the non-merging systems involve small MBHs which are leftovers from the seeding procedure. In future work we will use the model of Spinoso et al (in preparation) to explore the effect of seeding  in the population of MBHBs.\\

Regarding the merger ratio, events with $t_{\rm dyn}^{\rm BH}\,{<}\,t_{\rm H}$ display $\rm m_R\,{>}\,0.1$ at all redshifts. At stellar masses $\rm {>}\,10^{10}\, M_{\odot}$ we find cases with $t_{\rm dyn}^{\rm BH}\,{>}\,t_{\rm H}$ characterized by very low $\rm m_R$ (${\lesssim}\,0.01$). We checked that these events corresponds to minor mergers between massive galaxies and small galaxy companions ($\rm {<}\,10^{9}\, M_{\odot}$) whose host nuclear MBH rarely exceeds $\rm 10^{5}\, M_{\odot}$. 
At $\rm M_{stellar}^{Central}\,{<}\,10^{10}\, M_{\odot}$ is less common to have events with small merger ratios, especially at $z\,{<}\,2$. This is a natural consequence of the \texttt{Millennium} resolution as the minimum resolved stellar mass of satellite galaxies, $\rm {\sim}\,10^8\,M_{\odot}$, is comparable with the mass of the central galaxy for  $\rm M_{stellar}^{Central}\,{<}\,10^{10}\, M_{\odot}$ \citep[see Fig. B2 of][]{IzquierdoVillalba2019}.
Despite the large $\rm m_R$ of these events, $t_{\rm dyn}^{\rm BH}$ values are on average relatively large. This is caused by both the small mass of the black holes ($\rm {<}\,10^5 \, M_{\odot}$) and the large $r_0$ (${\sim}\, 10\, \rm kpc$) characterizing these events. Finally, the plane $\mathrm{M_{stellar}^{Central}}\,{-}\,t_{\rm dyn}^{\rm BH}$ seems to display a correlation with the galaxy morphology. In particular, the larger is the B/T the smaller is $t_{\rm dyn}^{\rm BH}$. This effect is particularly evident at $z\,{<}\,2$, where elliptical galaxies ($\rm B/T\,{>}\,0.7$) host pairing black holes with lower $t_{\rm dyn}^{\rm BH}$.\\

\subsection{Hardening and gravitational wave phase} \label{sec:HardeningModel}
As soon as the pairing phase ends\footnote{We assume that the pairing phase ends when $(t^{\rm Gal}_{\rm merge} \,{-}\, t_{\rm now})\,{-}\,t_{\rm dyn}^{\rm BH} \,{<}\,0$. $t^{\rm Gal}_{\rm merge}$ correspond to the lookback time at which the galaxy merger takes place and and $t_{\rm now}$ is the lookback time of the simulation.}, we assume that the MBHs form a hard binary with the central one. From hereafter, we tag as \textit{primary} black hole (with mass $\rm M_{BH,1}$)  the most massive black hole in the system whereas the less massive one is refereed as \textit{secondary} black hole (with mass $\rm M_{BH,2}$). The initial semi-major axis of the binary, $a_{\rm BH}$, is set to the scale in which $\mathrm{M_{Bulge}}(\,{<}\,a_{\rm BH}) \,{=}\,2\rm M_{BH,2}$, where $\mathrm{ M_{Bulge}}(\,{<}\,a_{\rm BH})$ is the mass in stars of the hosting bulge within $a_{\rm BH}$. In this work we assume that the evolution of the binary system depends on the type of environment in which is hosted. In particular, following  \cite{Antonini2015} we distinguish between two different type of environments that drive the two MBHs to  final coalescence: mergers in \textit{gas rich} and \textit{gas poor} environments. \\
 
Mergers in gas rich environments  require  the binary to be surrounded by a gas reservoir with a mass larger than the mass of the binary (i.e, $\rm  M_{\rm Res} \,{>}\, M_{\rm Bin}$, \citealt{Antonini2015}). In this case, the shrinking of the binary separation and the subsequent final coalescence is driven by the interaction with a massive circumbinary disk and gravitational wave emission. This scenario is supported by the results of the hydrodynamical simulations of \cite{Escala2004,Escala2005}, \cite{Dotti2007} and \cite{Cuadra2009} which showed that dense gaseous circumbinary disks are effective in shrinking MBHBs, promoting their coalescence in less than $\rm {\lesssim}\,10^7\,yr$ \citep[see also the work of][]{Armitage2002,Kocsis2012}. Given such effectiveness of the circumbinary gas disks in driving the MBHB to the final coalescence,  we neglect the stellar hardening effect. In this work we follow the results of \cite{Dotti2015} (see also \citealt{Bonetti2018a}) assuming that the evolution of the binary semi-major axis can be inferred from:
\begin{equation}\label{eq:BH_separation_GW_inspiral_gas}
\begin {split}
& \frac{d a_{\rm BH}}{dt} \,{=}\,  \left( \frac{d a_{\rm BH}}{dt}\right)_{\rm Gas} +  \left( \frac{d a_{\rm BH}}{dt}\right) _{\rm GW} \\
& \,{=}\, - \frac{2\dot{\mathrm{M}}_{\rm Bin}}{\mu} \sqrt{\frac{\delta}{1-e_{\rm BH}^2}} \, a_{\rm BH} - \frac{64G^3 (\mathrm{M}_{\rm BH_1}\,{+}\,\mathrm{M}_{\rm BH_2})^3 F(e_{\rm BH})}{5c^5(1+q)^2 a_{\rm BH}^3},
\end{split}
\end{equation} 
where the first and second term take into account the gas hardening and gravitational wave emission, respectively. Regarding the variables, $G$ is the gravitational constant, $c$ the light speed, $\delta\,{=}\,(1+q)(1+e_{\rm BH})$, $q\,{=}\,\rm M_{BH,2}/M_{BH,1}$, $\dot{\mathrm{M}}_{\rm Bin}$ is the sum of the accretion rate of both MBHs in the binary and $\mu$ is the reduced mass of the binary. Finally, $F(e)$ is a function which depends on the binary eccentricity \citep{PetersAndMathews1963}:
\begin{equation}
F(e_{\rm BH}) \,{=}\, (1-e_{\rm BH})^{-7/2}\left[ 1 + \left( \frac{73}{24} \right)e_{\rm BH}^2  + \left( \frac{37}{96} \right)e_{\rm BH}^4 \right] ,
\end{equation}

Here, we assume a fixed initial value of $e_{\rm BH}=0.6$ when the dynamics is gas-dominated and the binary is surrounded by a circum-binary disk (first term in Eq.~\ref{eq:BH_separation_GW_inspiral_gas}). This value is motivated by the work of \cite{Roedig2011} who found that the binary eccentricity coasts to a constant value of ${\sim}\,0.6$.  As soon as the GW emission (second term in Eq.~\ref{eq:BH_separation_GW_inspiral_gas}) dominates the MBHB evolution, we track the eccentricity evolution as \citep{Sesana2006}:

\begin{equation} \label{eq:eccEvolGas}
\begin{split}
\frac{de_{\rm BH}}{dt} \,{=}\, - \frac{304}{15} \frac{G^3 q (M_{\rm BH_1} + M_{\rm BH_2})^3}{c^5(1+q)^2 a_{\rm BH}^4 (1-e_{\rm BH}^2)^{5/2}} \left( e_{\rm BH} + \frac{121}{304}e_{\rm BH}^3\right) ,
\end{split}
\end{equation}

We highlight that if a binary system evolving in a gas rich environment exhausts the gas reservoir before the final coalescence, we switch to the equations describing the evolution in gas-poor environments, which we now provide.\\

For mergers in gas poor environments, we assume that the gas reservoir around the MBHs is smaller than the total mass of the binary (i.e, $\rm M_{\rm Res} \,{<}\, M_{\rm Bin}$). In this case, the hardening is caused by the extraction of binary energy and angular momentum through 3-body interactions with background stars that cross the  binary orbit  \citep{Quinlan1997,Sesana2006,Vasiliev2014,Sesana2015}. As for the gas rich case, the emission of gravitational wave starts to dominate when the hardening time becomes comparable to the gravitational wave timescale. In particular, in this type of environments the binary separation is tracked by integrating numerically the equation \citep{Sesana2015}:
\begin{equation}\label{eq:BH_separation_GW_inspiral}
\begin {split}
\frac{d a_{\rm BH}}{dt} \,{=}\, & \left( \frac{d a_{\rm BH}}{dt}\right)_{\rm Stars} +  \left( \frac{d a_{\rm BH}}{dt}\right) _{\rm GW} \\
& \,{=}\, -\frac{G H \rho_{\rm inf}}{\sigma_{\rm inf}} a_{\rm BH}^2 - \frac{64G^3 (\mathrm{M}_{\rm BH_1}\,{+}\,\mathrm{M}_{\rm BH_2})^3 F(e_{\rm BH})}{5c^5(1+q)^2 a_{\rm BH}^3},
\end{split}
\end{equation}
where $G$ is the gravitational constant, $c$ is the light speed and $H\approx 15-20$ is the hardening rate extracted from the tabulated values of \cite{Sesana2006}. The values of $\rho_{\rm inf}$ and $\sigma_{\rm inf}$ correspond respectively to the density and velocity dispersion of stars at the MBHB sphere influence. For this type of environments we assume that the binary systems start with an initial eccentricity randomly selected between $ 0 \,{<}\, e_{\rm BH} \,{<}\, 1$\footnote{We have tested the model by assuming that the initial eccentricity of the hardening phase is inherited from the one computed in the pairing phase (see Section~\ref{sec:PairingPhase}). We have found that such change leaves unaffected the stochastic GW background reported in this work.}. 
Besides, scattering experiments and numerical simulations in this type of environments indicate that the binary eccentricity is not constant during the hardening and GW phase but it changes through stellar encounters \citep{Hills1983,Mikkola1992,Quinlan1997,Sesana2006}. In particular, the variation of the eccentricity of the MBHB can be expressed as:
\begin{equation} \label{eq:eccEvol}
\begin{split}
& \frac{de_{\rm BH}}{dt} \,{=}\, \left( \frac{d e_{\rm BH}}{dt}\right)_{\rm Stars} +  \left( \frac{d e_{\rm BH}}{dt}\right) _{\rm GW} \\
& \,{=}\, a_{\rm BH}\frac{G \rho_{\rm inf} HK}{\sigma_{\rm inf}}  \\ & - \frac{304}{15} \frac{G^3 q (M_{\rm BH_1} + M_{\rm BH_2})^3}{c^5(1+q)^2 a_{\rm BH}^4 (1-e_{\rm BH}^2)^{5/2}} \left( e_{\rm BH} + \frac{121}{304}e_{\rm BH}^3\right),
\end{split}
\end{equation}
where $K$ is the eccentricity growth rate whose value is taken according to the Table 2 of \cite{Sesana2006}. \\

The values $r_{\rm inf}$, $\rho_{\rm inf}$ and $\sigma_{\rm inf}$ of Eq.~\ref{eq:BH_separation_GW_inspiral} and  Eq.~\ref{eq:eccEvol} were computed assuming a bulge mass profile. Unlike other works which use isothermal sphere or Dehnen profiles \citep[see e.g][]{Volonteri2003,Sesana2010,Sesana2015,Bonetti2018,Volonteri2020}, here we decided to use a Sérsic model. This choice is motivated by observational studies that found it to be a good approximation for fitting the bulge light distribution of different galaxies \citep{Drory2007,Drory2008,Gadotti2009}. The analytical expressions for the Sérsic model are taken from \cite{Prugniel1997} \citep[see also][]{Terzic2005}:
\begin{equation}\label{eq:Sersic}
\rm \rho (\mathit{r}) = \rho_0 \left( \frac{\mathit{r}}{R_e}\right)^{-p} \mathit{e}^{-b\left(\frac{\mathit{r}}{R_e}\right)^{1/n}},
\end{equation}

\begin{equation}\label{eq:Sigma_Sersic}
\begin{split}
\rm \sigma^2(\mathit{r})\,{=}\, & \frac{4 \pi G \rho_0^2 \mathrm{R_{e}}^2 n^2 b^{2\mathit{n}(p-1)}}{\rho(\mathit{r})} \\ & \int_Z^{\infty} \mathcal{Z}^{-\mathit{n}(p+1)-1} e^{-\mathcal{Z}} \gamma(\mathit{n}(3-p),\mathcal{Z}) d\mathcal{Z},
\end{split}
\end{equation}
where $\rm R_e$ is the bulge effective radius\footnote{We refer to \cite{Guo2011} for the explanation about the computation of bulge radius and \cite{IzquierdoVillalba2019} for improvements performed in the calculation of the bulge size after mergers.}, $\rho_0$ is the central bulge density, $n$ its Sérsic index. This index correlates with the central concentration of the bulge, being the bulges with smaller $n$, the ones less centrally concentrated. Finally, the variable $\gamma$ represents the incomplete gamma function, whereas $Z,p,\mathrm{b}$ are three different quantities that depend on the bulge properties: $\rm Z\,{=}\,b(\mathit{r}/R_e)^{1/n}$, $p \,{=} \, 1 - 0.6097 n^{-1} + 0.05563 n^{-2}$, $\mathrm{b}\,{=}\, 2n - 0.33 + 0.009876 n^{-1}$. This Sérsic model causes that smaller MBHs spend more time in the hardening phase than the most massive ones. 
To guide the reader, for a MBHB system with total mass $\rm M_{bin}\,{=}\,10^9\,M_{\odot}$, $q\,{=}\,1$ and $e_{\rm BH}\,{=}\,0.3$, the hardening time-scale is $\rm {\sim}\,0.2\, Gyr$. For the same system but with $\rm M_{bin}\,{=}\,10^6\,M_{\odot}$, the time increases up to $\rm 10\, Gyr$. For further details we refer to \cite{Biava2019} where a detailed study of hardening time scales in different bulge profiles was performed. \\

\begin{figure}
\centering
\includegraphics[width=1.\columnwidth]{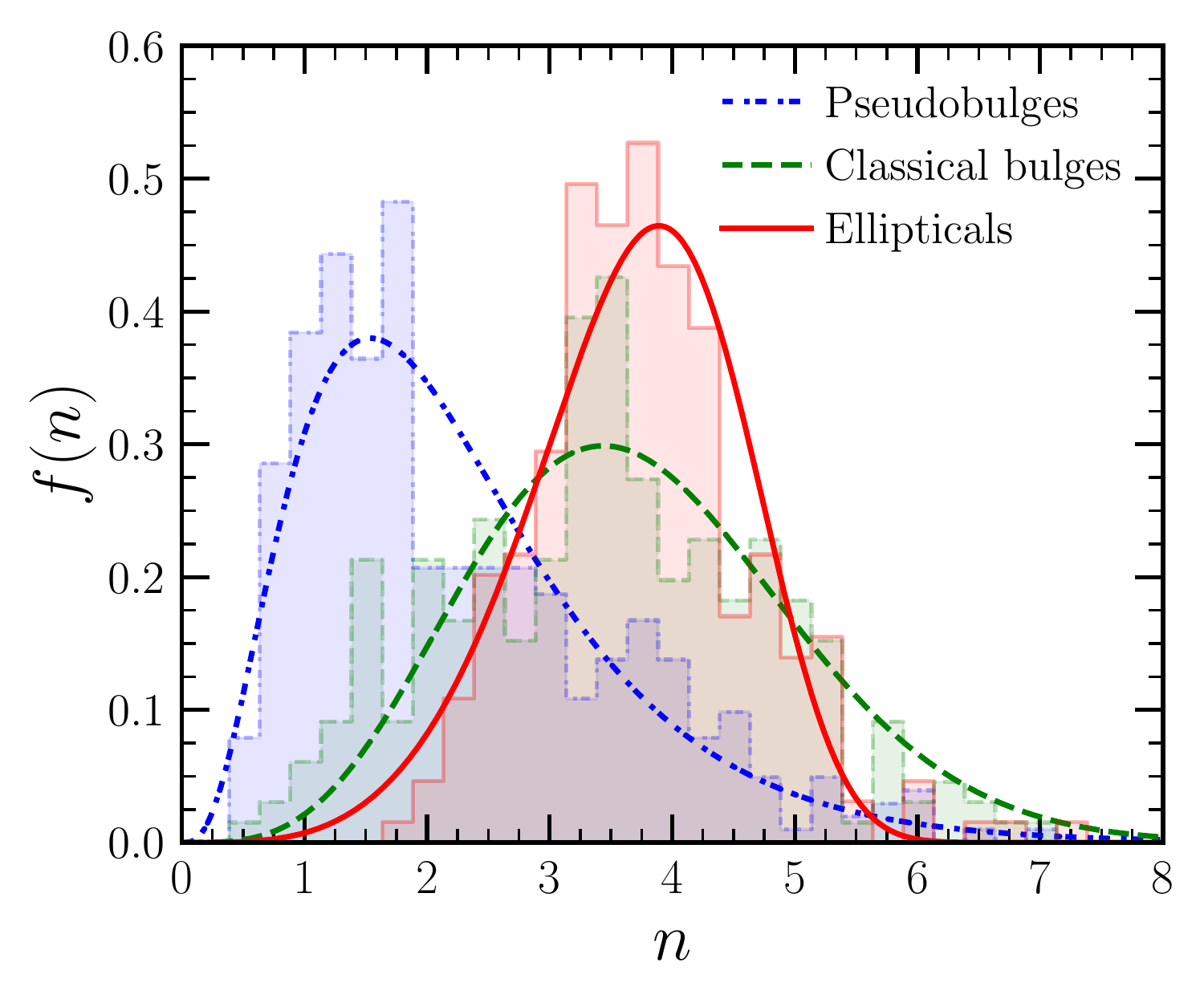}
\caption[]{The histograms display the Sérsic index, $n$, distribution extracted from \protect{\cite{Gadotti2009}}: Elliptical structures, classical bulges and pseudo-bulges are displayed in red (solid line), green (dashed line) and blue (dotted-dashed line), respectively. The solid lines display the fits to these histograms according to \protect{Eq.~\ref{eq:BulgeFit}}. 
The Sérsic index correlates with the central concentration of the bulge. In particular, the smaller is $n$, the less centrally concentrated is the bulge.}
\label{fig:FitSersicIndex}
\end{figure}

One of the disadvantages of \LGalaxies is that it does not compute Sérsic indexes, but only the mass assembled throughout different channels of growth: major and minor mergers assemble elliptical and classical bulges, whereas disk instabilities prompt pseudobulges. To attach a Sérsic value to each galaxy, we compute the Sérsic index distribution of $z\,{=}\,0$ pseudobulges, classical bulges and elliptical galaxies  using the observational data provided by \cite{Gadotti2009}. For each bulge type, we  fit their distributions according to:
\begin{equation}\label{eq:BulgeFit}
    f(n) = A\left( \frac{n}{n_0} \right) e^{-(n/n_0)^{\beta}},
\end{equation}
where $A$, $n_0$ and $\beta$ are  free parameters. In Fig.~\ref{fig:FitSersicIndex} we show the fits for pseudobulges, classical bulges and elliptical galaxies. Table~\ref{Table:FitsSersic} contains the best fit for these parameters. As we can see, each bulge type follows a distinct distribution, and the larger differences are seen between pseudobulges and elliptical galaxies. Such difference have been reported in the last years, highlighting that the formation scenario of each bulge type might leave an imprint in the stellar dynamics and distribution \citep{Kormendy1983,Kormendy1996,Drory2007,Drory2008,Elmegreen2008,Gadotti2009}. Once determined the Sérsic index distribution, the way of assigning these values to  \LGalaxies bulges is as follows: each time a galaxy develops/increments the bulge via DI (minor, major merger), we extract a Sérsic index from the pseudobulge (classical bulge, elliptical) fit. If the galaxy had an already existing bulge, the final Sérsic index is computed as the mass-weighted average of the old bulge and the extra mass added to it. We highlight that the observations of \cite{Gadotti2009} only take into account galaxies with stellar mass $\rm {>}\,10^{10} \, M_{\odot}$, removing from the sample dwarf galaxies. In this work, we assume that the fits presented in Table~\ref{Table:FitsSersic} hold at any stellar mass. We further assume that the $z\,{=}\,0$ Sérsic indexes distribution of pseudobulges, classical bulges and ellipticals  hold at higher redshifts. This is a simplification and such values might evolve in the real Universe. Nevertheless, the results of \cite{Shibuya2015} suggest that starforming galaxies do not display a redshift evolution in their median Sérsic index ($n\,{\sim}\,1.5$).\\ 

\begin{table}
	\centering
	\setlength{\tabcolsep}{10pt}
	\renewcommand{\arraystretch}{0.8}
	\begin{tabular}{cccc}
		\hline
		Bulge type & $A$ & $n_0$ & $\beta$  \\ \hline \hline
		\textit{Elliptical} &  $1.15\,{\pm}\,0.04$ &  $4.24\,{\pm}\,0.09$ & $5.75\,{\pm}\,0.71$\\ 
		\textit{Classical bulge} &  $0.60\,{\pm}\,0.13$ & $2.47\,{\pm}\,0.41$ & $1.88\,{\pm}\,0.31$ \\
		\textit{Pseudobulge} &  $0.021\,{\pm}\,0.005$ &  $0.166\,{\pm}\,0.08$ & $0.71\,{\pm}\,0.09$\\ \hline
	\end{tabular}
	\caption{Parameters for elliptical, classical- and pseudo- bulges from Eq.~\ref{eq:BulgeFit}.}
	\label{Table:FitsSersic}
\end{table}

\subsection{Black hole triplets in galactic nuclei}

As we discussed in the previous section, the lifetime of a binary system at the center of a galaxy is fully determined by the hardening phase. However, in some instances the efficiency of this process in shrinking the MBHB separation down to the GW phase can be very low \citep{Milosavljevic2001,Yu2002,Merritt2005,Sesana2007}. Indeed, if the hardening time scale is long enough, a third black hole in the pairing phase can reach the galaxy center and interact with the MBHB system \citep{Hoffman2007,Kulkarni2012}. If this happens, the interaction between the three MBHs can lead to the prompt coalescence of two of them or a scattering event (usually ejecting the lighter MBH). Indeed, \cite{Bonetti2018ModelTriplets} demonstrated that these interactions are a plausible mechanism for triggering a merger in stalled binaries. In this work we treat the triple black hole interaction by including in \LGalaxies the model of \citealt{Bonetti2018ModelGrid}. In particular, we use the \citealt{Bonetti2018ModelGrid} tabulated values to select those triple interactions which lead to the merger of a pair of MBHs and those causing the ejection of the lighter MBH from the system. In this latter case the separation of the leftover MBHB is computed following \cite{Volonteri2003} and the final eccentricity is select as a random value between $[0\,{-}\,1]$. This grid model of \cite{Bonetti2018ModelTriplets} needs as an input three values: the mass of the primary black hole, the binary mass ratio and $\rm M_{BH,1}/(M_{BH,2}\,{+}\,M_{BH,3})$ (where $\rm M_{BH,3}$ is the mass of the intruder black hole).

\subsection{The growth of pairing black holes and hard binaries} \label{sec:GrowthPairingHardening}

The recent hydrodynamical simulations of  merging galaxies with central MBHs by \cite{Capelo2015} showed that the secondary galaxy suffers large perturbations during the pericenter passages around the central one. In these circumstances  the black hole of the secondary galaxy experiences accretion enhancements, mainly correlated with the galaxy mass ratio. In this work we include these findings assuming that right before the galaxy merger, the black hole of the secondary galaxy is able to generate or increase its gas reservoir. 
In this work we determine the amount of mass deposited in the MBH reservoir according to Eq.\ref{eq:QuasarMode_Merger}. The growth in this pairing phase is modelled in the same way as we did for nuclear black holes, i.e the accretion rate is determined by an initial Eddington limited phase followed by a self-regulated growth in which the black hole consumes the gas at low Eddington rates \citep[see][for the equations that govern that growth phase]{IzquierdoVillalba2020}.\\

Gas accretion onto MBHB systems has been extensively studied during the last years \citep{DOrazio2013,Farris2014,Moody2019,Munoz2019,DOrazio2021}. Despite not being a simple process to study and model, it has been possible to draw a general picture. The circumbinary disk gas is progressively stripped from its inner edges, feeding trough accretion streams mini-disk around the two MBHs which ultimately are accreted. Interestingly, it has been shown that irrespective of the mass ratio of the binaries, the gas accretion onto the secondary black hole is sufficient to change the final mass ratio of the binary, moving the initial values toward larger ones \citep[see e.g][]{Farris2014,Duffell2020}. Based on this picture, during the hardening phase of the MBHB system we follow the results of \cite{Duffell2020}. Accordingly,   the accretion rate of a primary black hole ($\dot{\rm M}_{\rm BH_1}$) is fully determined by the binary  mass ratio ($q$) and the accretion rate of the secondary black hole ($\dot{\rm M}_{\rm BH_2}$):
\begin{equation} \label{eq:Relation_accretion_hard_binary_blac_hole}
\dot{\rm M}_{\rm BH_1} =  \dot{\rm M}_{\rm BH_2} (0.1+0.9\mathit{q}),
\end{equation}
Therefore, each time a hard binary has formed surrounded by an circumbinary accretion disk, we fix the accretion of the secondary black hole at the Eddington limit and we determine the accretion onto the primary based on Eq.~\ref{eq:Relation_accretion_hard_binary_blac_hole}.

\section{Results} \label{sec:Results}
In this section we present the main results. 
We infer from \LGalaxies the  chirp mass distribution of the MBHBs and merger rates from the model. We then report on the predictions for the amplitude of the gravitational wave (GW) background at the ${\sim}\,\rm nHz$ frequencies proved by the PTA experiments. Finally, we generate two variants of the model where the GW background amplitude is increased by pushing the gas accretion onto the MBHs after  mergers and disk instabilities (see Eq.~\ref{eq:QuasarMode_Merger} and Eq.~\ref{eq:QuasarMode_DI}). We explored the capability to produce a population of massive black holes compatible with current constraints from observational works. The results on the amplitude of the GW stochastic background is tested against current knowledge on the AGN and MBH mass distributions recalling that the model of BH growth and spin evolution of \LGalaxies has been calibrated to be consistent with this set of observations \citep[see][]{IzquierdoVillalba2020}.

\subsection{Merged black holes: Chirp masses and merger ratios}
\begin{figure}
\centering
\includegraphics[width=1.\columnwidth]{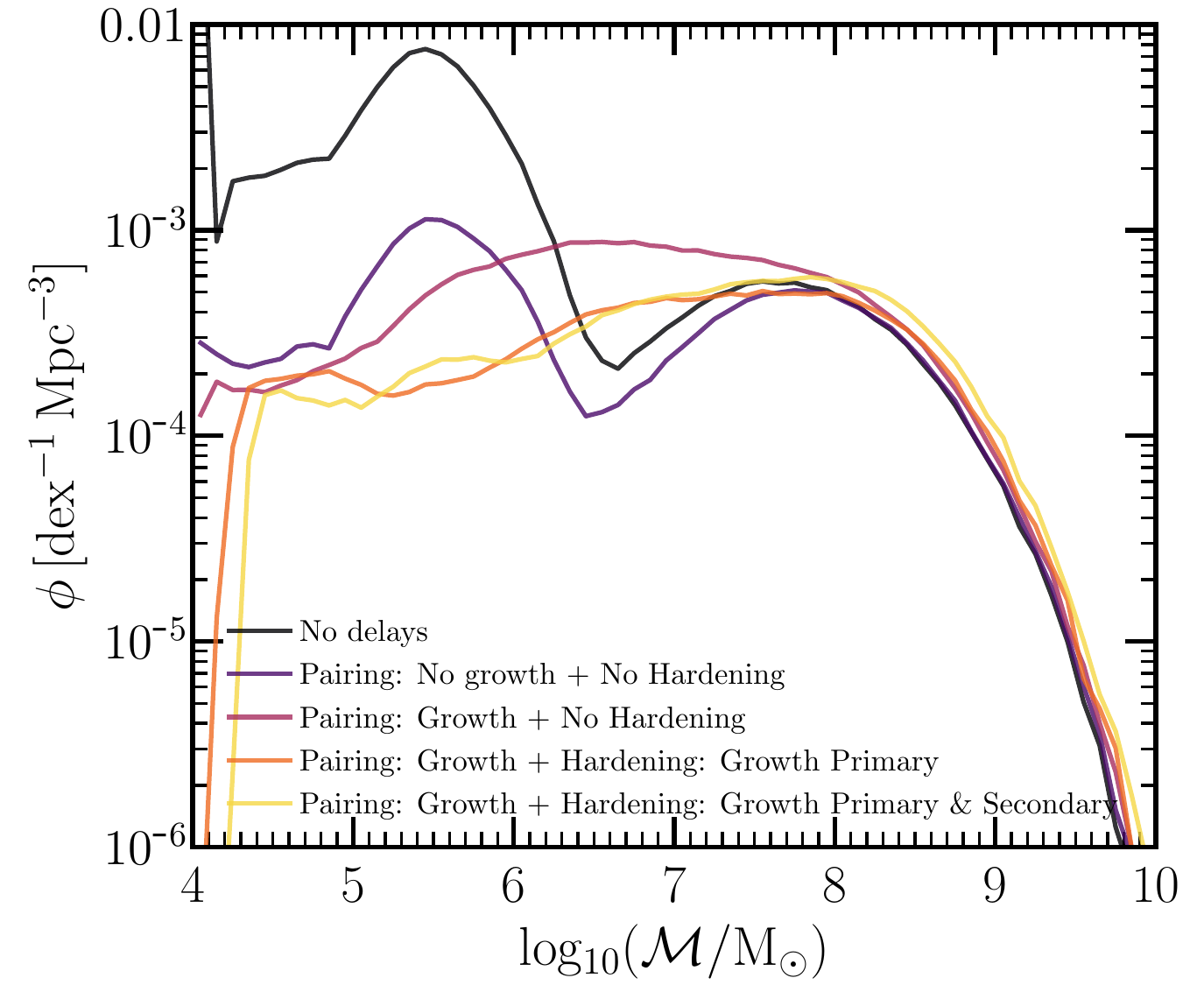}
\caption[]{Chirp mass function of the \textit{merged} black holes. Black line refers to the case  when no delays are assumed in the model. Violet and purple lines represent the chirp mass function with only the  delay  in the pairing phase with and without gas accretion onto the in-spiralling black holes, respectively. Orange  (yellow) line represents the model with both pairing and hardening phase with the secondary black hole in the binary  able (unable) to accrete matter form the circum-binary disk.}
\label{fig:ChirpMassFunctions}
\end{figure}

The chirp mass of a binary, in the source frame,  is the quantity that takes an important role in the amplitude of the GW emitted by a coalescing binary and is defined as:
\begin{equation} \label{eq:chirp}
{\mathcal{M}}= \rm \frac{(M_{\rm BH,1}M_{\rm BH,2})^{3/5}}{(M_{\rm BH,1}+M_{BH,2})^{1/5}}
\end{equation}
In Fig.~\ref{fig:ChirpMassFunctions} we present the \textit{rest-frame} chirp mass function of merged black holes. When no delays are included in the model, we can see a large population of mergers with $\mathcal{M}\,{<}\,10^{6} \, \rm M_{\odot}$. This is an artifact produced by the seeding model, where all the newly resolved galaxies are seeded with a fix $10^4\, \rm M_{\odot}$ seed black hole. Nevertheless, when a dynamical friction time-delay is added in the pairing phase (without any hardening phase), the merger rate of low-mass binary systems is reduced. In particular, we can see that below $\mathcal{M}\,{\lesssim}\,10^6\, \rm M_{\odot}$  the mass function has decreased by a $\rm {>}\,1 \, dex$. Interestingly, the pairing phase does not have an effect on the high mass end of the distribution, where no significant differences are found. This is  caused by the fact that MBHs of $\rm M_{BH}>10^7\,M_{\odot}$ have a short pairing time scale, typically ${<}\,100\rm \, Myr$ (see Fig.\ref{fig:t_dyn_BH_tidal_All_MS}).\\

In the same Fig.~\ref{fig:ChirpMassFunctions} we explored the effects of black hole mass-growth during the pairing phase. We refer the reader to Section~\ref{sec:GrowthPairingHardening} for the treatment used to deal with MBH growth in the pairing phase. Notice that, as we did before, no hardening is added yet. Therefore as soon as the pairing phase is over, a MBH merger takes place. As shown in the figure, the main difference between the model with and without growth is that the former gives a larger number of events at $\mathcal{M}\,{>}\,10^{6} \, \rm M_{\odot}$. This different behavior is caused by the effectiveness of the growth during the pairing phase in reducing the mass difference between the pairing black hole and nuclear MBH at the time of the binary formation and its subsequent coalescence \citep{Capelo2015}. Interestingly, the larger differences are found at  $\mathcal{M}\,{<}\,10^{7.5}\,\rm M_{\odot}$ which arise from the fact that secondary MBHs involved in these mergers display $\rm M_{BH}{<}\,10^{6}\, \rm M_{\odot}$ and $t_{\rm dyn}^{\rm BH}\,{\lesssim}\,1\, \rm Gyr$. Such large time delays allow these MBHs to consume all (or most of) the gas reservoir stored during the pre-merger phase. On the contrary, at $\mathcal{M}\,{>}\,10^{7.5}\,\rm M_{\odot}$, the secondary MBHs ($\rm M_{BH}{>}\,10^{6}\, \rm M_{\odot}$) display $t_{\rm dyn}^{\rm BH}\,{\lesssim}\,0.1\, \rm Gyr$, having less time to increase their masses before the coalescence. When the hardening phase is included on top of the pairing one, the chirp mass function changes principally at $\mathcal{M}\,{<}\,10^{7}\, \rm M_{\odot}$ where the amplitude decreases a factor ${\sim}\,4$. On the other had, the massive end is almost untouched. This different mass behaviour is just the natural consequence of the evolution of hard binaries in Sérsic model profiles. As discussed in Section~\ref{sec:HardeningModel}, the larger is the mass of the binary system the smaller is the hardening time scale \citep[see][]{Biava2019}. Particularly, MBHB systems with total mass $\rm M_{bin}\,{=}\,10^9\,M_{\odot}$ display a hardening time-scale $\rm {\sim}\,0.2\, Gyr$, whereas for $\rm M_{bin}\,{=}\,10^6\,M_{\odot}$ the time increases up to $\rm {\sim}\,10\, Gyr$. Thus, the decay of the mass function at $\mathcal{M}\,{<}\,10^{7}\, \rm M_{\odot}$ is the effect of the MBHBs stalling at the hardening phase.\\

The hardening phase explored before only allows accretion onto the primary black hole during the lifetime of the MBHB system. 
However, as discussed in Section~\ref{sec:GrowthPairingHardening}, we included the possibility of the secondary MBH to accrete matter from the cirbumbinary disk which surrounds the binary system. In Fig.~\ref{fig:ChirpMassFunctions} we present the chirp mass function for that case. As shown, no big differences are seen at $\mathcal{M}\,{<}\,10^{8}\, \rm M_{\odot}$ when we compare the hardening model with and without the growth of the secondary MBHs.  The larger differences are displayed in the massive end ($\mathcal{M}\,{>}\,10^{8}\, \rm M_{\odot}$), where there is a clear increase of the mass function for the hard model with gas accretion onto the secondary MBH. This effect has been also seen in some recent works based on the post-processing of hydrodynamics simulations. For instance, \cite{Siwek2020} found that boosting the growth of the secondary black hole over the primary one causes a shift of the chirp mass function towards large masses.\\

\begin{figure}
\centering
\includegraphics[width=1.\columnwidth]{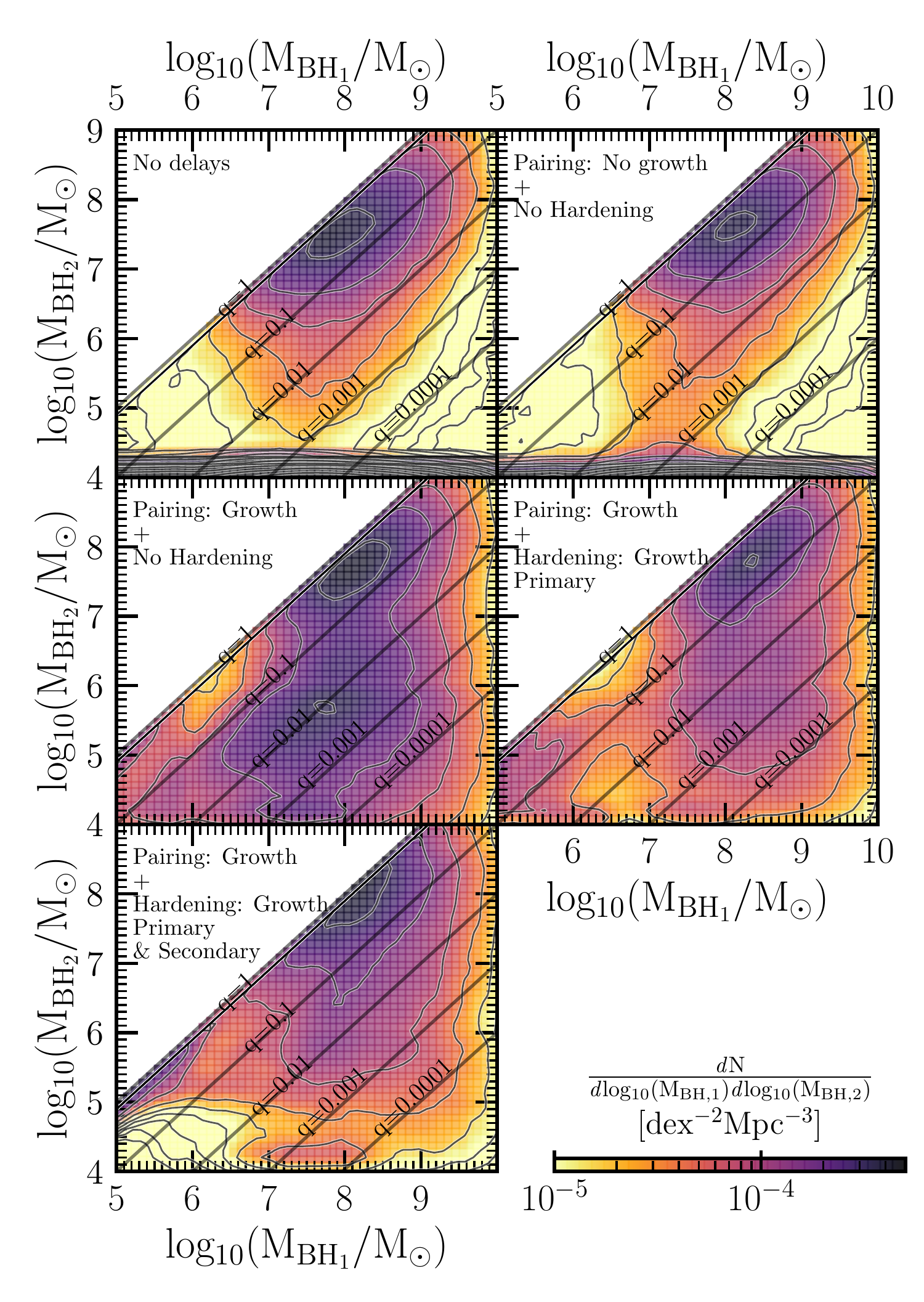}
\caption[]{Relation between the mass of the primary and secondary black hole ($\rm M_{BH_{1}}$ and $\rm M_{BH_{2}}$, respectively). Different black lines highlight different binary mass ratios, $q\,{=}\,1,0.1,10^{-2},10^{-3},10^{-4}$. \textbf{Upper left panel}: Predictions when no delays in the black hole mergers are assumed. \textbf{Upper right panel}: Delay due to the pairing phase. \textbf{Middle left panel}: Accretion of the remaining accretion disk is allowed during the pairing phase. \textbf{Middle right panel}: Delay by both pairing and hardening phase. The primary black hole consumes the whole circumbinary disk. \textbf{Lower left panel}: Delay by both pairing and hardening phase. The secondary black hole is able to accrete part of the circumbinary disk.}
\label{fig:q_MBH1_MBH2}
\end{figure}

In Fig.~\ref{fig:q_MBH1_MBH2} we analyze the effect of different delays and gas accretion prescriptions on the distribution of merging binaries in the $\rm (M_{BH,1},M_{BH,2})$ plane. In the first panel we present the results when no delays are added. For $\rm M_{BH,1}\,{>}\,10^5\,\rm M_{\odot}$ a large number of mergers happen with seed mass black holes, causing that the most of the primary MBHs ($\rm M_{BH,1}\,{>}\,10^7\,\rm M_{\odot}$) display merger ratios $q\,{<}\,10^{-3}$ (dark horizontal black stripe at the bottom of the panel). Despite this, the models finds a significant number of events with $\rm M_{BH,2}>M_{seed}$ and $q\,{>}\,0.01$ although a large scatter is seen, especially at $\rm M_{BH,1}\,{<}\,10^8\, M_{\odot}$. The second panel of the Fig.~\ref{fig:q_MBH1_MBH2} presents the same but when the pairing phase is added. No big differences are see, except a large decrease of the mergers involving seed mass MBHs (see Fig.~\ref{fig:ChirpMassFunctions} to see better such drop). 
When we add the growth in the pairing phase (third panel of Fig.~\ref{fig:q_MBH1_MBH2}), we see significant changes. In particular at $ \rm 10^7\,{<}\,M_{BH,1}\,{<}\,10^8 \, M_{\odot}$ the mergers happen with more massive secondary black holes. In this mass range, the mass of the involved secondary MBH displays a bi-modality. There is a big cloud at $ \rm 10^5\,{<}\,M_{BH,2}\,{<}\,10^6 \, M_{\odot}$, which prompt mergers with $0.001{<}\,q\,{<}\,0.1$. As we already discussed, such secondary MBHs increased their final $q$ values at the coalescence time thanks to their large pairing times ($t_{\rm dyn}^{\rm BH}\,{\lesssim}\,1\, \rm Gyr$) which allow them to consume most of the gas reservoir stored during the pre-merger phase. On the other hand, we can see a secondary cloud at $ \rm 10^{7.5}\,{<}\,M_{BH,2}\,{<}\,10^8 \, M_{\odot}$. Although it was already present in the pairing model without growth, in this case the number of events has increased. Although the typical merger ratios ($q\,{>}\,0.1$) are more shifted towards $q\,{=}\,1$, no large differences are seen with respect to the ones of the pairing phase without growth. As commented before, the small pairing times of these secondary MBHs ($t_{\rm dyn}^{\rm BH}\,{\lesssim}\,0.1\, \rm Gyr$) disfavor large mass changes during the pairing phase. When a hardening phase is added (fourth panel of Fig.~\ref{fig:q_MBH1_MBH2}), a large number of mergers with $ \rm 10^5\,{<}\,M_{BH,2}\,{<}\,10^6 \, M_{\odot}$ vanishes. In this case, the merger ratios that predominate are the ones $q\,{>}\,0.1$. Finally, when we allow the growth of the secondary black hole during the hardening phase (fifth panel of Fig.~\ref{fig:q_MBH1_MBH2}) we see an effect of systematically increasing the $q$ parameter regardless the value of $\rm M_{BH,1}$. Indeed, in this case most of the mergers with $\rm M_{BH}\,{>}\,10^8\,M_{\odot}$ have $q\,{\sim}\,1$. As discussed before, this effect is also seen by \cite{Siwek2020} which exploring different growth models found that rising the mass accreted by the secondary black hole, causes an increase of black hole merger events close to $ q\,{=}\,1$. \\

From here on, we will consider our \textit{fiducial model} to be the one
in which growth is allowed in both pairing and hardening. 
Specifically, during the hardening phase, we allow both \textit{primary} and \textit{secondary} MBH to accrete matter from the circumbinary disk.\\

\subsection{The gravitational wave stochastic background}
\begin{figure}
\centering
\includegraphics[width=1.\columnwidth]{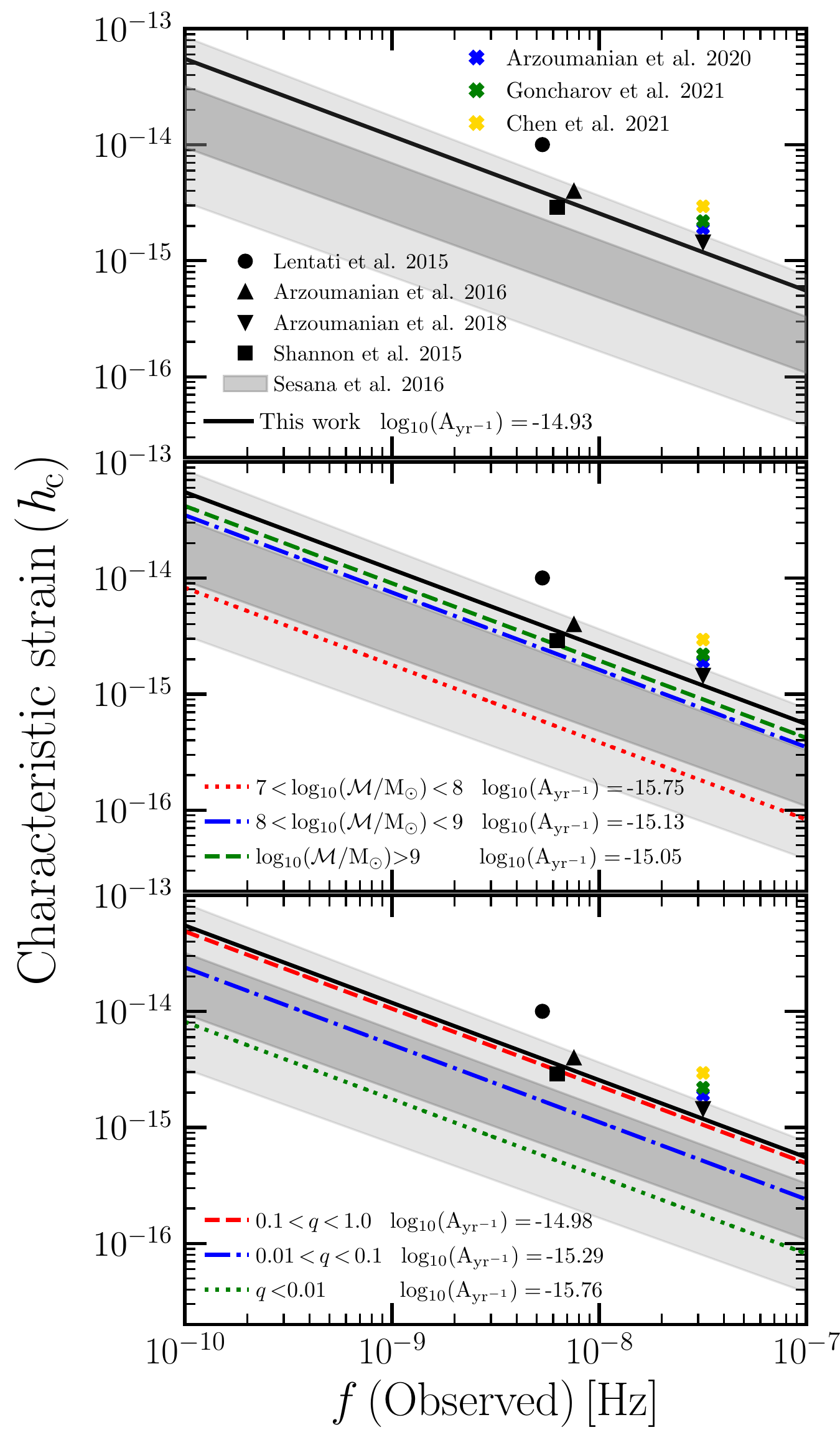}
\caption[]{\textbf{Upper panel}: Gravitational wave background amplitude predicted by the model. \textbf{Middle panel}: Gravitational wave background signal computed in three different chirp mass intervals: $\rm 10^{7} \,{<}\, \mathcal{M}\,{<}\,10^{8}\, M_{\odot}$ (red dotted line), $\rm 10^{8} \,{<}\, \mathcal{M}\,{<}\,10^{9}\, M_{\odot}$ (blue dotted-dashed line) and $\rm \mathcal{M}\,{>}\,10^{9}\, M_{\odot}$ (green dashed line). \textbf{Lower panel}: Gravitational wave background signal same as above split in three binary mass ration: $0.1\,{<}\,q\,{<}\,1.0$ (red dashed line), $0.01\,{<}\,q\,{<}\,0.01$ (blue dotted-dashed line) and $q\,{<}\,0.01$ (green dotted line). In all the three plots the circle, triangle and square points are the upper limits placed by the EPTA \protect \citep{Lentati2015}, the NANOGrav \citep{Arzoumanian2016,Arzoumanian2018} and the PPTA \protect \citep{Shannon2015} projects, respectively. Blue, green and yellow crosses at $f{(\rm Observed)}\,{=}\,  \rm 1yr^{-1}$ represents, respectively, the measurements of the common red noise reported by \protect \cite{Arzoumanian2020}, \protect \cite{Goncharov2021} and \protect \cite{Chen2021}. The shaded areas show the constrains coming from the local Universe bulge-black hole relation \protect \citep{Sesana2016GWB}: dark and clear grey areas represent the $1$ and $2$ $\sigma$ confidence interval.}
\label{fig:GWBackground}
\end{figure}

Following \cite{Sesana2008}, the characteristic stochastic gravitational wave background from a population of inspiralling MBHBs can be expressed as:
\begin{equation} \label{eq:GWB1}
    h^2_c(f) \, {=} \, \frac{4 G^{5/3}}{f^{2} c^2\pi} {\int}{\int} \frac{dz d\mathcal{M}}{(1+z)} \frac{d^2n}{dzd\mathcal{M}}\frac{d\mathrm{E}_{\rm GW}(\mathcal{M})}{d\ln{f_r}},
\end{equation}
where $d^2n/dzd\mathcal{M}$ is the comoving number density of MBHB merger per unit redshift, $z$, and rest-frame chirp mass, $\mathcal{M}$ and $f$ is the frequency of the GWs in the observer frame. The quantity $d\mathrm{E_{GW}}/d\ln{f_r}$ represents the differential energy spectrum of the binary, i.e the energy emitted per logarithmic rest-frame frequency, $f_r$. Given that we are interested in the population of inspiral MBHB in the PTA band, we make the specific assumption that the MBHBs producing the GWB are in perfect circular orbits evolving purely due to GW emission. 
From these assumptions, Eq.~\ref{eq:GWB1} can be re-written as:
\begin{equation} \label{eq:GWBG2}
    h^2_c(f) \, {=} \, \frac{4 G^{5/3} f^{-4/3}}{3c^2\pi^{1/3}} {\int}{\int} dz d\mathcal{M} \frac{d^2n}{dzd\mathcal{M}} \frac{\mathcal{M}^{5/3}}{(1+z)^{1/3}},
\end{equation}
which is often \textbf{expressed} as:
\begin{equation}
h_c(f) \,{=}\, A\left( \frac{f}{f_0} \right)^{-2/3} 
\end{equation}
where $A$ is the amplitude of the signal at the reference frequency $f_0$. Usually, the gravitational wave background amplitude is referred at $f_0 \,{=}\ \rm 1yr^{-1}$. Hereafter, we will denote $\mathrm{A}(f_0\,{=}\,1 \rm yr^{-1})$ as $A_{\rm yr^{-1}}$. In Fig.~\ref{fig:GWBackground} we present the model predictions. The value of $A_{\rm yr^{-1}}$ corresponds to ${\sim}\,1.2{\times}10^{-15}$, being in agreement with the upper limits placed by the EPTA \citep{Lentati2015}, the NANOGrav \citep{Arzoumanian2018} the PPTA \citep{Shannon2015} and the IPTA \citep{Verbiest2016} projects. The model is also compatible with the predictions coming from the bulge-black hole relation in the local Universe \citep{Sesana2016GWB}. Other works based on semi-analytical models or hydrodynamics simulations displayed similar results. For instance, \cite{Kelley2016} by using the \texttt{Illustris} simulation, reported $A_{\rm yr^{-1}}\,{=}\,7.1\,{\times}10^{-16}$. Despite the good agreement with other works, Fig.~\ref{fig:GWBackground} shows that our predictions are below the most recent results of  NANOGrav (12.5-year data analysis, \citealt{Arzoumanian2020}), PPTA (DR2, \citealt{Goncharov2021}) and EPTA (DR2, \citealt{Chen2021}). Section~\ref{sec:PushingGWB} will be devoted to the comparison between our predictions and NANOGrav/PPTA/EPTA latest results, trying to reconcile theoretical results with observational constraints. \\ 


\begin{figure}
\centering
\includegraphics[width=1.\columnwidth]{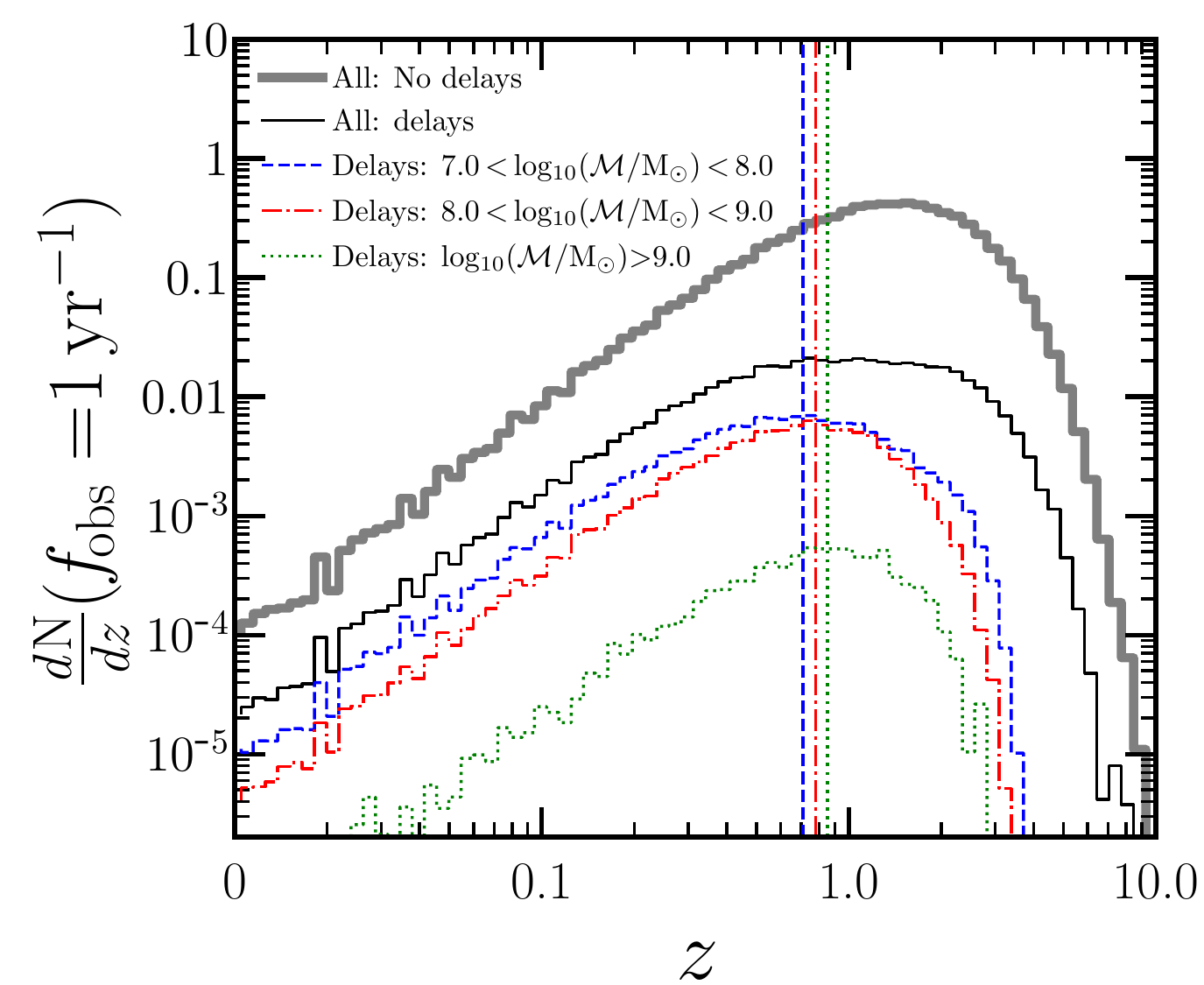}
\caption[]{Merger rates predicted by \LGalaxies. The black thick line displays the predictions of the model when no delays (pairing and hardening) are assumed. Think black thick line represents the same but with pairing and hardening delays. Colored lines represent the merger ratios in the model with delays at different chirp masses: $ \rm 7\,{<}\, log_{10}(\mathcal{M}/M_{\odot}) \, {<}\, 8$ (blue dashed line), $ \rm 8\,{<}\, log_{10}(\mathcal{M}/M_{\odot}) \, {<}\, 9$ (red dotted-dashed line) and $ \rm log_{10}(\mathcal{M}/M_{\odot}) \, {>}\, 9$ (green dotted line). The vertical lines highlight the maximum of each distribution.}
\label{fig:MergerRate}
\end{figure}

In the middle panel of Fig.~\ref{fig:GWBackground}, we show the GW spectrum signal produced by binary systems of three different chirp masses: $\rm 10^{7} \,{<}\, \mathcal{M}\,{<}\,10^{8}\, M_{\odot}$, $\rm 10^{8} \,{<}\, \mathcal{M}\,{<}\,10^{9}\, M_{\odot}$ and $\rm \mathcal{M}\,{>}\,10^{9}\, M_{\odot}$. As shown, the two latter bins contribute the most to the signal. On the other extreme, binaries of $\rm 10^{7} \,{<} \mathcal{M}\,{<}\,10^{8}\, M_{\odot}$ have a marginal effect, contributing typically $\rm 1 \, dex$ less than $\rm \mathcal{M}\,{>}\,10^{8}\, M_{\odot}$. Regarding the mass ratios of MBHBs generating the GW background, the bottom panel of Fig.~\ref{fig:GWBackground} shows that systems with $q\,{>}\,0.1$ are the ones producing most of the signal. Furthermore, the results show that the smaller the $q$ parameter, the smaller is the effect of the binary system in the GW background. For instance, binary systems with $0.01\,{<}\,q\,{<}0.1$ and $q\,{<}\,0.01$ generate respectively 0.4 and 0.13 times smaller amplitude than the total signal. These results are consistent with \cite{Sesana2008} and \cite{Sesana2013} which showed that ${>}\,95\%$ of the GW signal at $\,{\sim}\, \rm nHz$ frequencies comes from BH major mergers ($q\,{>}\,0.25$) involving BHs with mass $\rm {>}\,10^8 \, M_{\odot}$ at $z\,{<}\,1.5$. Similar results were recently reported by \cite{CaseyClyde2021}. By using empirical relations for quasar luminosity functions, quasar lifetime and MBHB mass ratio distribution, the authors concluded that most of the GWB signal 
would be produced by MBHBs of mass $\rm {>}\,10^8 \, M_{\odot}$ at $z\,{\sim}\,0.5$. \\

In Fig.~\ref{fig:MergerRate} we present the merger rates for MBHB without any binary treatment (thin black line) and when we included the pairing and hardening delay (think black line). The figure shows that the MBHB model causes a large change in the rates at which the MBHs coalesce. Whereas the integrated merger rate without MBH merger delays reaches up to $\rm 1.03 \, yr^{-1}$, in the version with delays it drops down to $\rm 0.06 \, yr^{-1}$. For the latter case,  we have explored the predictions for $\mathcal{M}\,{>}\,10^7 \, \rm M_{\odot}$. As we can see, the mergers of these massive binaries happen at relatively low-$z$, being typically at $z\,{\sim}\,1$. When the population is divided into different mass bins, a mild redshift difference is seen, being the systems with larger $\mathcal{M}$ the ones that merge slightly earlier. Besides, at $z\,{<}\,1$ merger events of $\mathcal{M}\,{>}\,10^9 \, \rm M_{\odot}$ binaries decrease faster than the ones of $\rm 10^{7} \,{<}\, \mathcal{M}\,{<}\,10^{8}\, M_{\odot}$ and $\rm 10^{8} \,{<}\, \mathcal{M}\,{<}\,10^{9}\, M_{\odot}$ which have similar behaviour.


\subsection{The stochastic gravitational background confronting the mass and quasar luminosity functions} \label{sec:PushingGWB}

\begin{figure}
\centering
\includegraphics[width=1.\columnwidth]{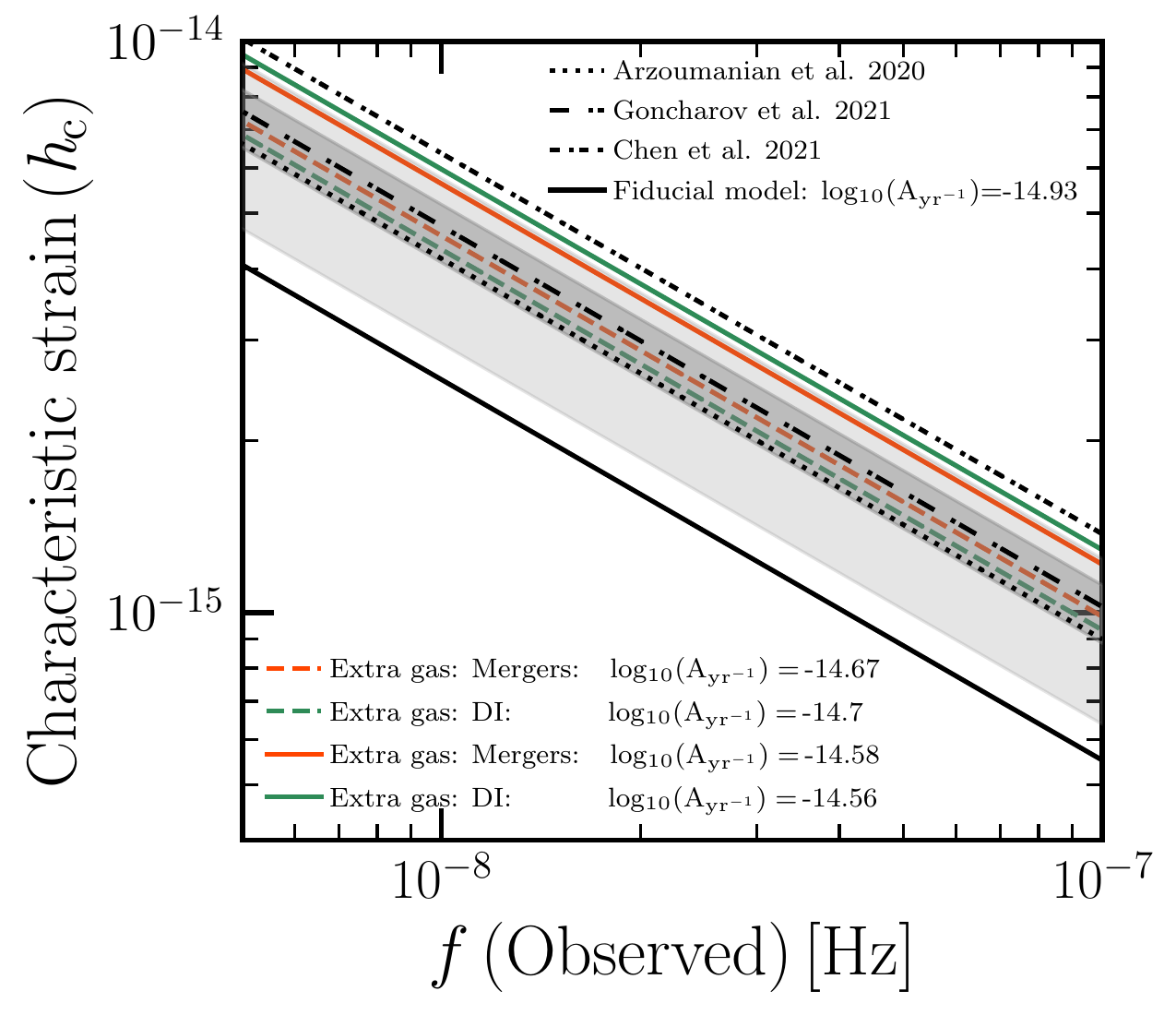}
\caption[]{Amplitude of the gravitational wave background in the frequency range $10^{-10}\,{-}\,10^{-7}\, \rm Hz$. While the solid black line represents the fiducial model, orange and green lines display the results when the gas accretion during merger and disk instabilities is boosted, respectively. Dashed (solid) lines represent the model predictions when $A_{\rm yr^{-1}} \,{\sim}\,1.92\,{\times}\,10^{-15}$ ($A_{\rm yr^{-1}} \,{\sim}\,2.67\,{\times}\,10^{-15}$) is reached. The clear grey shaded area and the dotted line show the constrains coming from \protect \cite{Arzoumanian2020}. The dark grey shaded area and the long dashed-dotted line show the constrains coming from \protect \cite{Goncharov2021}. Short dashed-dotted line represents the results of \protect \cite{Chen2021} (to avoid confusion we did not show the upper and lower limits of  \protect \citealt{Chen2021}). }
\label{fig:GWBBoosted}
\end{figure}


Recently, by using the $12.5$-yr pulsar-timing data set of NANOGrav collaboration \cite{Arzoumanian2020} reported strong evidences of a stochastic process with $A_{\rm yr^{-1}}$ spanning between $1.37\,{\times}\,10^{-15} \,{-}\, 2.67\,{\times}\,10^{-15}$ (5\%-95\% quantiles) and median value of  $1.92\,{\times}\,10^{-15}$. Similar signal was also recently reported by the PPTA ($A_{\rm yr^{-1}} \,{\sim}\,1.9\,{\times}\,10^{-15} \,{-}\, 2.6\,{\times}\,10^{-15}$) and EPTA ($A_{\rm yr^{-1}} \,{\sim}\, 2.23\,{\times}\,10^{-15} \,{-}\, 3.8\,{\times}\,10^{-15}$) second data release \citep{Goncharov2021,Chen2021}. Even though such signals did not display significant evidences of quadrupolar correlations needed to claim GW detection, it is interesting to test which are the predictions of our model for such large GW signal. Specifically, in this section we present the model predictions when reaching a GWB compatible with the median value and 95\% quantiles of \cite{Arzoumanian2020}. The conclusions presented in this section are the same when the limits of \cite{Goncharov2021} (PPTA) and \cite{Chen2021} (EPTA) are used. \\


To increase the GW signal we explored two variants of the model. The first one consisted in increasing the amount of gas accreted by the black holes during galaxy mergers (hereafter model \textit{increased merger}, IM) by increasing the parameter $\rm \mathit{f}_{BH}^{\rm merger}$ (see Eq.~\ref{eq:QuasarMode_Merger}) by a factor of $2$ and $3$ to reach $A_{\rm yr^{-1}} \,{\sim}\,1.92\,{\times}\,10^{-15}$ and  $A_{\rm yr^{-1}} \,{\sim}\,2.67\,{\times}\,10^{-15}$, respectively. In the second variant of the model we left the mergers untouched and changed the gas accretion during disk instabilities (hereafter model \textit{increased DI}, IDI). Specifically, we increase $\rm \mathit{f}_{BH}^{\rm DI}$ (see Eq.~\ref{eq:QuasarMode_DI}) by a factor of $9$ and $20$ to achieve respectively a GWB of $A_{\rm yr^{-1}} \,{\sim}\,1.92\,{\times}\,10^{-15}$ and  $A_{\rm yr^{-1}} \,{\sim}\,2.67\,{\times}\,10^{-15}$. 
The GW backgrounds produced by these four model variants are presented in Fig.~\ref{fig:GWBBoosted}.\\

\begin{figure*}
\centering
\includegraphics[width=1.75\columnwidth]{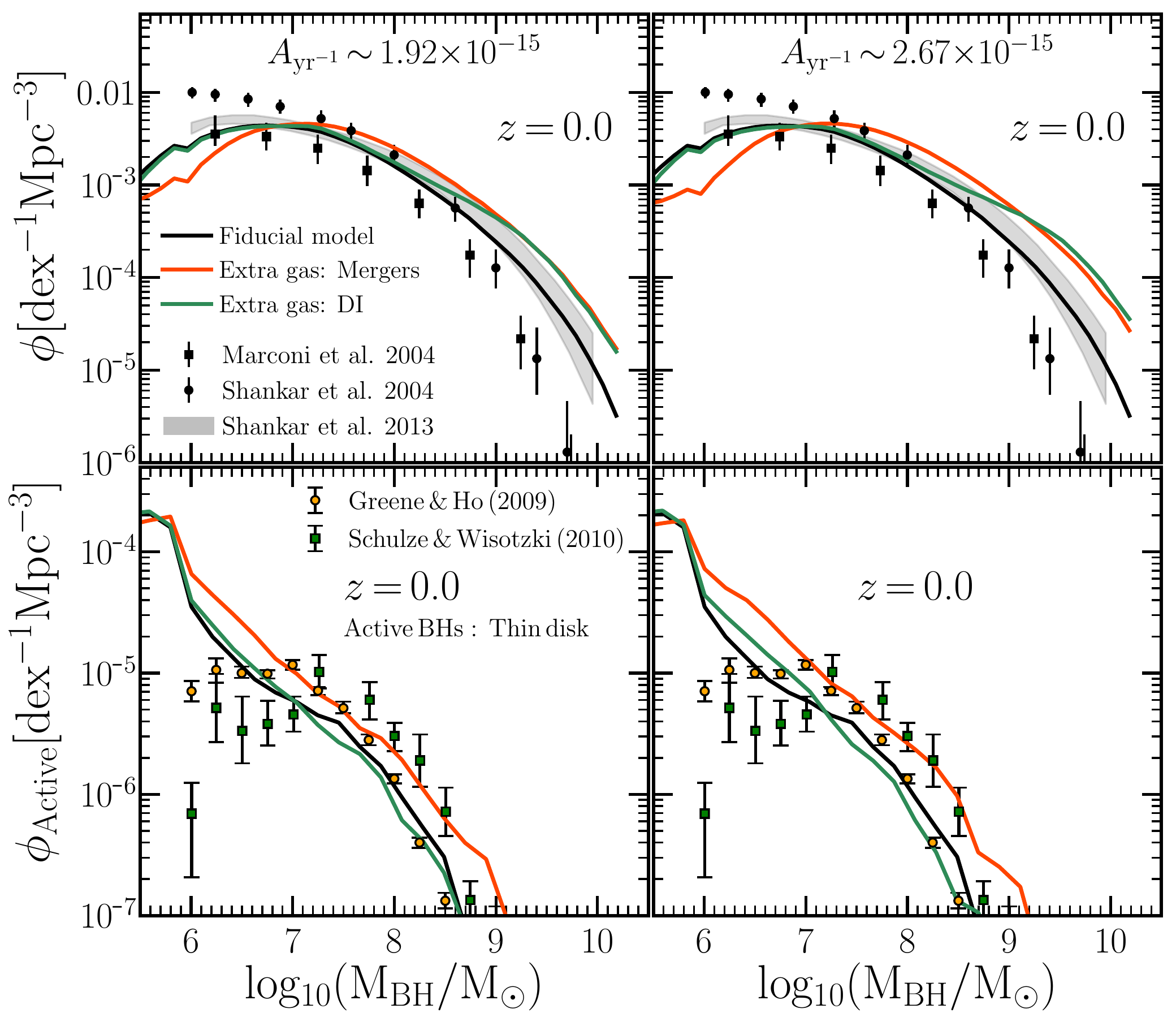}
\caption[]{Model predictions when a stochastic GWB of $A_{\rm yr^{-1}}\,{\sim}\,1.92\,{\times}\,10^{-15}$ (left panels) and  $A_{\rm yr^{-1}}\,{\sim}\,2.67\,{\times}\,10^{-15}$ (right panels) is reached. The upper panel display the $z\,{=}\,0$ black hole mass function compared to the observational results of \protect \cite{Marconi2004} and \protect \cite{Shankar2004,Shankar2013}. The lower panels show the black hole mass function at $z\,{\sim}\,0$ for active black holes (Eddington ratio $\,{>}\,{10^{-2}}$) from \protect\cite{Greene2007} and \protect \cite{Schulze2010} are added for comparison. In all the plots the black line corresponds to the predictions of the fiducial model. Orange and green lines represent the results when we boost the gas accretion during mergers and disk instabilities, respectively. }
\label{fig:BHMF_Boosted}
\end{figure*}

The question to answer now is whether these new models are also  consistent with constraints on the black hole mass and luminosity function. 
In Fig.~\ref{fig:BHMF_Boosted} we present the comparison between the models and the current observations of the black hole mass function (BHMF) in the local Universe \citep{Marconi2004,Shankar2004,Shankar2009,Shankar2013}. As shown, our fiducial run is in good agreement with these observations. On the other hand, the models with a boosted mass growth display values in tension with the observations, especially the ones with GWB of $A_{\rm yr^{-1}}\,{\sim}2.67\,{\times}\,10^{-15}$. Regardless of the GWB level, in the IDI cases, we see a behaviour compatible with observations for $\rm 10^6\,{<}\,M_{BH}\,{<}\,10^8 \, M_{\odot}$. However, the massive end ($\rm M_{BH}\,{>}\,10^8 \, M_{\odot}$) is typically over-predicted by almost a factor $3$ for $A_{\rm yr^{-1}}\,{\sim}\,1.92\,{\times}\,10^{-15}$ and ${\sim}\,1\,\rm dex$ for $A_{\rm yr^{-1}}\,{\sim}\,2.67\,{\times}\,10^{-15}$. A similar trend is observed in the IM models. Additionally, the latter show incompatibilities at lower masses as well ($\rm M_{BH}\,{\sim}\,10^{6.5}\, M_{\odot}$). Even though the high mass tail of the BHFM seems to be not fully constrained by observations and there is still room for further improvements, some authors pointed out that current MBH mass estimates might be biased high. For instance, as reported by \cite{Shankar2016}, discrepancy between \cite{Shankar2013} and \cite{Marconi2004} might be caused by biases affecting the observations. \cite{Shankar2016} argues that, because of selection effects, the normalization of the scaling relations used to relate the black hole mass with galaxy properties (such as bulge mass and velocity dispersion) might be increased by a factor as high as ${\gtrsim}\,3$ \citep[see also][]{Bernardi2007,Shankar2019}. Therefore, this will yield a lower amplitude in the empirical relations which would cause smaller measurements of BH masses and BH mass density, consistent with the current non-detection of this signal by pulsar timing array experiments \citep[see][]{Sesana2016}.\\

In the lower panels of Fig.~\ref{fig:BHMF_Boosted} we show the mass function of active MBHs, selected as those with Eddington ratios larger than $0.01$. The predictions are compared with \cite{Greene2007} and \cite{Schulze2010} which performed the same Eddington ratio selection. As shown, regardless the GWB amplitude, the fiducial, IM and IDI models are consistent with the predictions at $\rm 10^{7}\,{<}\,M_{BH}\,{<}\,10^8 \, M_{\odot}$. However, the IM models over-predict the population of active MBHs at $\rm M_{BH}\,{>}\,10^{8.5}$. 
For masses $\rm {<}\,10^{7}\, M_{\odot}$ we can not draw strong conclusions when comparing predictions with observations, considering current selection effects of the latter. For instance, the flux limit imposed by \cite{Schulze2010} causes large incompleteness effects at at low black hole masses and low Eddington ratios.\\

\begin{figure*}
\centering
\includegraphics[width=1.\columnwidth]{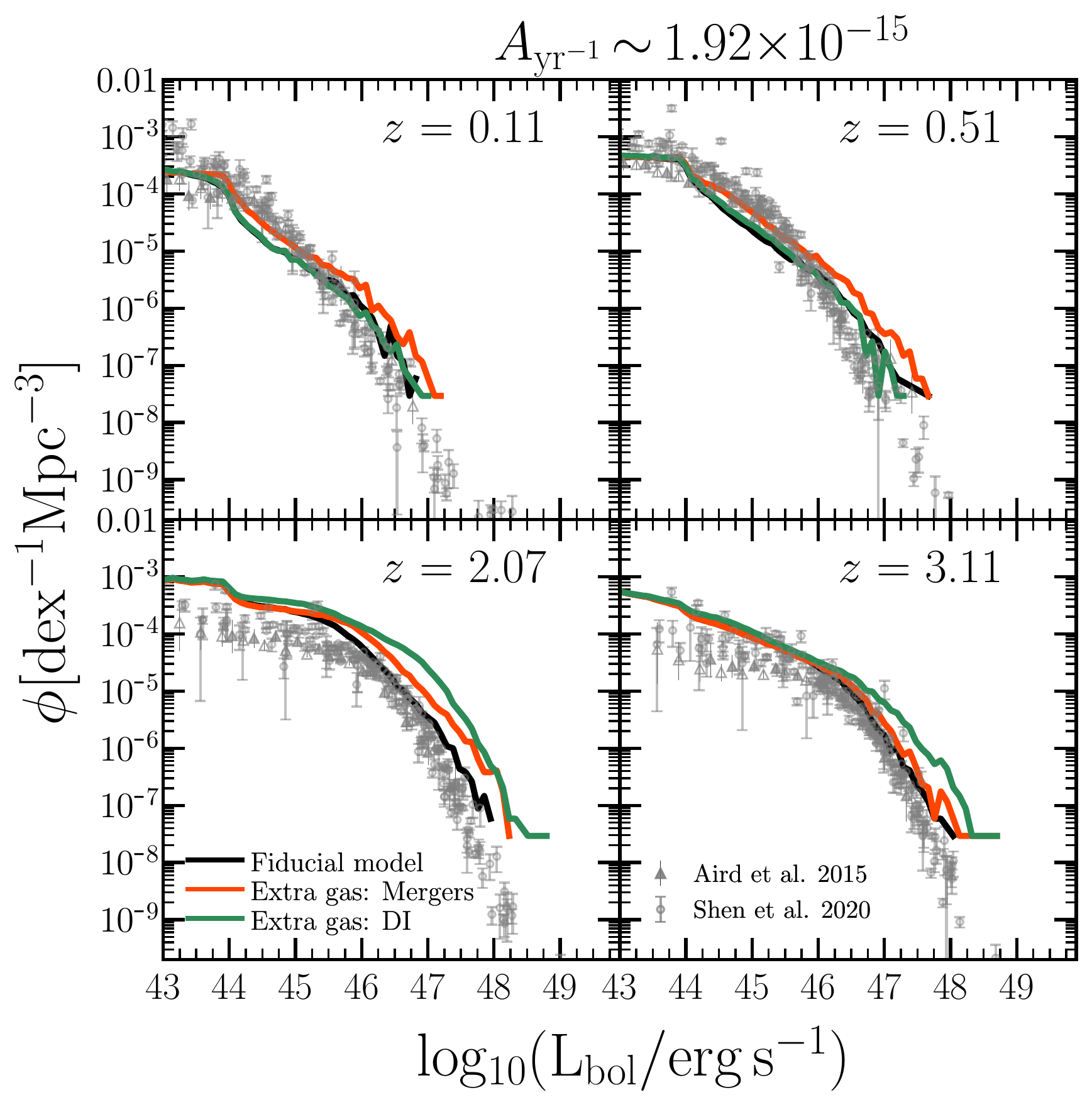}
\includegraphics[width=1.\columnwidth]{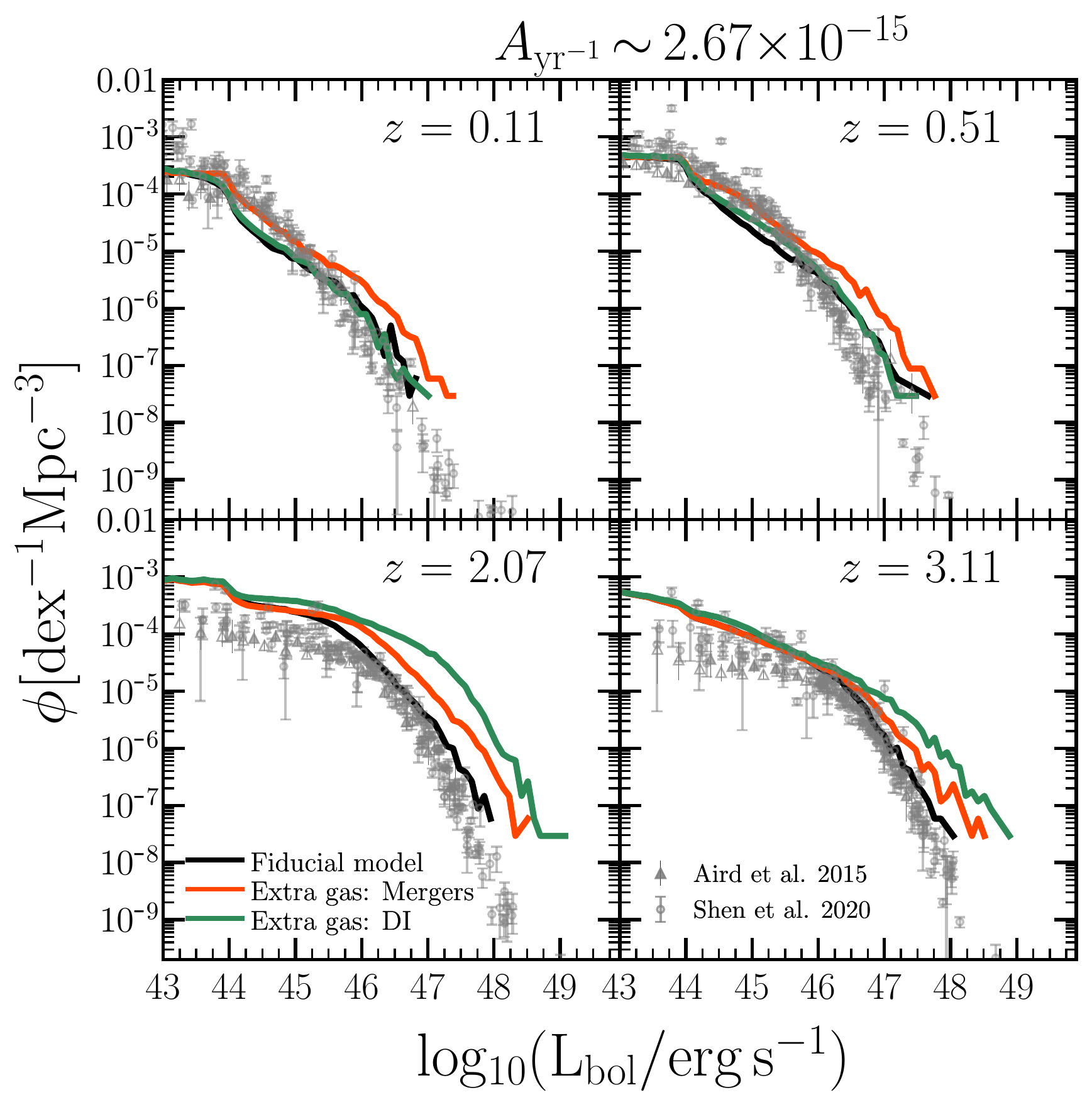}
\caption[]{Quasar bolometric luminosity functions ($\rm L_{bol}$) at $z\,{\approx}\,0.1, 0.5, 2.0, 3.0$. Luminosity functions are compared with the data of \protect\cite{Shen2020} (circles) and \protect\cite{Aird2015} (triangles). The left (right) panels correspond to the model predictions when a stochastic GW background of $A_{\rm yr^{-1}}\,{\sim}\,1.92\,{\times}\,10^{-15}$ ($A_{\rm yr^{-1}}\,{\sim}\,2.67\,{\times}\,10^{-15}$) is reached. In all the plots the black line corresponds to the predictions of the fiducial model. Orange and green lines represent the results when we rise the gas accretion during mergers and disk instabilities, respectively.}
\label{fig:QLF_Boosted}
\end{figure*}

In Fig.~\ref{fig:QLF_Boosted} we present the evolution of the quasar bolometric luminosity function (LF) from $z\,{\sim}\,3$ down $z\,{\sim}\,0$. Even though these functions give the number density of accreting black holes in different luminosity bins, they have been a powerful tool to extract information on how massive black holes grow with cosmic time, on the geometry of the accretion disks and other fundamental quantities such as the black hole spins and radiative efficiencies. In this work we only focus on the very bright objects, i.e ${>}\,10^{45}\, \rm erg/s$, avoiding the comparison with lower luminosity given the current limitations on observational and theoretical models. In particular, from an observational standpoint, the  covered area and depth of current surveys pose serious challenges when extracting statistical properties of the  LF at the faint end \citep{Siana2008,Masters2012,McGreer2013,Niida2016,Akiyama2018}. Even more, dust attenuation effects might play an important role in shaping current measurements. On the other hand,  current theoretical works show  a large excesses at luminosity ${<}\,10^{45}\, \rm erg/s$. In order to reconcile observations with predictions these works have played with empirical relations for obscuring accreting black holes or  with the efficiency of the seeding process \cite[see e.g][]{Degraf2010,Fanidakis2012,DeGraf2020}. Even though these works provide interesting results shedding light on the nature of low-luminous quasars, the treatment of seeding or dust obscuration is beyond the scope of this paper. As shown in Fig.~\ref{fig:QLF_Boosted} the fiducial model is compatible with current observations of the quasar LF, showing a sharp cut off at larger luminosity \citep{Shen2020}. On the other hand, the models with higher gas accretion display a completely different behavior. Boosting the gas accretion during DI leads to a larger excess of bright quasars at $z\,{>}\,1.0$. For instance, at $z\,{\sim}\,2$ and for luminosities ${>}\,10^{46}\, \rm erg/s$ the models with $A_{\rm yr^{-1}} \,{\sim}\,1.92\,{\times}\,10^{-15}$ and  $A_{\rm yr^{-1}} \,{\sim}\,2.67\,{\times}\,10^{-15}$ are systematically over-predicting the number density by a factor ${\sim}\,1\,\rm dex$ and ${\sim}\,2\,\rm dex$, respectively. A similar behavior is seen at $z\,{\sim}\,3$. At lower redshifts ($z\,{<}\,1.0$) the model follows both the fiducial results and the observed trends. This is principally caused by the decrease of important disk instabilities events at these redshifts. Regarding the IM models, we can see similar trends at $z\,{>}\,2$, where the bright end of the LF is systematically larger than the observed one. We highlight that the difference is larger with $A_{\rm yr^{-1}} \,{\sim}\,2.67\,{\times}\,10^{-15}$. Interestingly, the excess with respect to the observations is smaller than with the IDI model. This is principally caused by the fact that DI events are more important than mergers at these redshifts  \citep{IzquierdoVillalba2020}. At lower redshifts, we can see larger differences with respect to the fiducial and the IDI models: IM model is systematically overprotecting the bright end of the LF (${>}\,10^{46}\, \rm erg/s$). Such differences can be a factor of $3$ ($1.5$) by $z\,{\sim}\,0$ up to a factor $5$ ($2$) at $z\,{\sim}\,0.5$ for $A_{\rm yr^{-1}} \,{\sim}\,1.92\,{\times}\,10^{-15}$ ($A_{\rm yr^{-1}} \,{\sim}\,2.67\,{\times}\,10^{-15}$).\\

Based on the results presented in Fig.~\ref{fig:BHMF_Boosted} and Fig.~\ref{fig:QLF_Boosted} we can draw the conclusion that large gravitational wave backgrounds can be reached by our semi-analytical model just by changing the gas accretion of the black holes after mergers or disk instabilities. However, these amplitudes are difficult to  reconcile  with  observational constrains such as the black hole mass function or quasar bolometric luminosity function. Therefore, we highlight that the reliability of GW backgrounds produced by both semi-analytical models or hydrodynamical simulations must be tested by checking the properties of the full black hole population such as luminosity functions or mass distribution across cosmic time. On this line, we can find the recent work of \cite{CaseyClyde2021} in which it is presented a new model to constrain the population of MBHB based on GW backgrounds and quasar populations. According to the number density of quasars and their expected lifetime \citep{Hopkins2006b,Hopkins2007} the authors pointed out that the last NANOGrav GW signal would suggest a local number density of MBHB $5$ times larger than the previously detected, being 25\% of the MBHB system associated with quasars.\\



\section{Summary and Conclusions}

In this paper we presented a model tracking the formation and evolution of massive black holes binaries (MBHBs) across cosmic time. We made use of the \LGalaxies{} semi-analytical model \citep[SAM,][]{Henriques2015} run on the \texttt{Millennium} dark matter merger trees whose mass resolution allows to draw solid conclusions for galaxies of mass ${\gtrsim}\,10^{8.5}\, \rm M_{\odot}$ and MBHs ${\gtrsim}\,10^{6}\,\rm M_{\odot}$. The MBHB model was developed as an extension of the work presented in \cite{IzquierdoVillalba2020} where detailed prescriptions for the mass growth and spin evolution of MBHs were included in \LGalaxies. In a nutshell, the MBHs are allowed to grow trough cold gas accretion, hot gas accretion and mergers with other black holes. Specifically, the former channel is the main driver of the black hole growth and it is triggered by both galaxy mergers and disk instabilities (DI). During any growth events, the code tracks the evolution of the black hole spin in a self-consistent way.   \\


Following the standard scenario, we included three different stages for the dynamical evolution of MBHBs that needs to be tracked in order to build a population of MBHBs: \textit{pairing}, \textit{hardening} and \textit{gravitational wave} (GW) phase. We assumed that the first phase starts after the galaxy-galaxy merger is completed, and corresponds to the sinking process of the MBH of the satellite galaxy towards the center of the newly formed  galaxy. The process is driven by  dynamical friction acting on the black holes individually, and exerted by the galaxy's stellar component. The time spent by the MBH of the less massive galaxy during the pairing phase has been computed following recent refinements of the \cite{Chandrasekhar1943} formula, which account for the eccentricity of the MBH orbit. Since the dynamical friction timescale depends on the initial position of the MBH relative to the host galaxy, this distance has been computed accounting for mass stripping of the secondary by the tidal field of the primary galaxy. 
The model has shown that the orbit of a large fraction of MBHs stalls in this phase, being a bottleneck for the formation of a bound MBHB system. Despite that, the number of MBHs with $\rm {>}\,10^6\,M_{\odot}$ reaching the galaxy nucleus increases towards low-$z$. On top of this, we have found that elliptical galaxies at $z\,{<}\,1.0$ are the preferred birthplaces of MBHB systems.\\ 

During the pairing phase, we  allowed the black holes to accrete their pre-merger gas reservoir. Interestingly, this has an imprint on the final chirp mass function (CBHMF) of merged MBHs. The main effects are seen at $10^{6}\,{<}\,\mathcal{M}\,{<}\,10^{7.5}\,\rm M_{\odot}$ where the CBHMF amplitude increases with respect to the case in which gas accretion is suppressed. Such change is due to the long-lived  phase of  dynamical friction (${\lesssim}\,1\, \rm Gyr$) experienced by the MBHs in these mergers. This has led to a significant increase of their masses, by consuming all the gas reservoir stored during the pre-merger phase. A similar trend is seen at $\mathcal{M}\,{>}\,10^{7.5}\,\rm M_{\odot}$, but the effects are smaller given the shorter timescales  involved in these cases, which disfavored large mass increases during the pairing phase.\\

When the pairing phase has ended, the MBHs form a binary system governed by the hardening and gravitational wave phase.
We distinguished between two different environments in this phase: \textit{gas rich} and \textit{gas poor}. In the former case, a circumbinary gas disk around the MBHB forms and  dominates the system. The torques exerted by the disk cause the shrinking of the orbit and coalescence of the two MBHs. In this  environment, the binary separation is tracked by integrating numerically the differential equation of \cite{Dotti2015}. By contrast, in gas poor environments, the hardening phase is caused by the effect of stars intersecting the MBHB orbit. These interactions are able to extract a significant amount of the MBHB energy and angular momentum through the slingshot mechanism. The binary separation and eccentricity in this type of environment are tracked by integrating numerically the differential equation of \cite{Sesana2015}, assuming a Sérsic model profile for the host galaxy \citep{Sersic1968}. Following the findings of \cite{Duffell2020}, we assumed that gas accretion during the hardening phase is determined by the binary mass ratio and the accretion rate of the secondary black hole, set to the Eddington limit. 
Finally, regardless of the environment, we included the \cite{Bonetti2018ModelTriplets} model for triplet reaction among a binary and an incoming black hole as an additional mechanism capable of driving stalled binaries to coalescence. The results show that binary hardening in gas poor environments reduces  significantly the number of MBHB merges at $\mathcal{M}\,{<}\,10^{7}\, \rm M_{\odot}$ while leaving untouched the high mass end of the CBHMF. This different mass behavior is caused by the evolution of hard binaries in Sérsic model profiles, where the lighter MBHBs have hardening times $\rm {\sim}\, 2 \, dex$ larger than the most massive ones.\\ 

Thanks to the large volume and mass resolution of the \texttt{Millennium} simulation, we explored the model to predict the amplitude of  the stochastic gravitational wave background (GWB) at the frequencies proved by the \textit{Pulsar timing array} (PTA) experiments. The model shows an amplitude at $1\, \rm yr^{-1}$ of $A_{\rm yr^{-1}}\,{\sim}1.2\,{\times}\,10^{-15}$, being principally produced by binary systems with $\rm \mathcal{M}\,{>}\,10^{8}\, M_{\odot}$ and $q\,{>}\,0.1$. The GWB reported in this work is in agreement with current upper limits provided by \cite{Lentati2015}, \cite{Arzoumanian2016} and \cite{Shannon2015} but in tension with the last constraints reported by \cite{Arzoumanian2020}. 
Therefore, we considered the amplitude identified by \cite{Arzoumanian2020} (under the hypothesis that is a GWB coming form MBHBs)
and asked what modifications to the model could produce a GWB level consistent with \cite{Arzoumanian2020} results. 
Only by boosting the MBH gas accretion during mergers and disk instabilities we produced a larger GW background amplitude ($A_{\rm yr^{-1}}\, {=} \, 1.37\,{\times}\,10^{-15} \,{-}\, 2.67\,{\times}\,10^{-15}$) more consistent with the amplitude recently reported by the NANOGrav collaboration \citep{Arzoumanian2020}. Unlike previous studies in the literature, we confronted the predictions on the amplitude of the stochastic GWB with constraints from key observations such as the quasar luminosity functions (LFs) and local black hole mass function (BHMF). In particular, large GW amplitude values ($A_{\rm yr^{-1}}\,{>}\,1.92\,{\times}\,10^{-15}$) made difficult to reconcile model predictions with the observational constraints. In particular, we showed that the models with large GWB display a large excess of bright quasars at any redshift. For instance, at $z\,{\sim}\,2$ quasars with luminosity ${>}\,10^{46}\, \rm erg/s$ are systematically over-predicted by a factor $2\,\rm dex$. At $z\,{<}\,0.5$, such over-prediction is still present, especially in the model where gas accretion onto mergers was boosted. Regarding the BHMF, the models with GWBs compatible with \cite{Arzoumanian2020} constraints display values in tension with the observations, especially in the massive end ($\rm M_{BH}\,{>}\,10^8 \, M_{\odot}$) where the difference with current observational constraints reach up to $1\,{-}\,1.5\,\rm dex$.\\

The model presented here is a step forward for the study of MBHBs across cosmic time. In future, thanks to the flexibility of the model, we will extend the analysis to the \texttt{MillenniumII} dark matter (DM) merger trees \citep{Springel2005,Boylan-Kolchin2009}. Their different box sizes and DM mass resolutions will offer the capability to explore the physical processes ruling the evolution of MBHs over a wider range of masses and environments. Therefore, we will be able to characterize  not only the formation, evolution, and environments of the most massive binary systems accessible through PTA experiments \citep{Kramer2013,McLaughlin2013,Manchester2013} but also the less massive ones proved by the \textit{Laser Interferometer Space Antenna} \citep[LISA,][]{LISA2017}.

\section*{Acknowledgements}
D.I.V and A.S acknowledge financial support provided under the European Union’s H2020 ERC Consolidator Grant ``Binary Massive Black Hole Astrophysics'' (B Massive, Grant Agreement: 818691). D.I.V. acknowledges also financial support from INFN H45J18000450006. M.C.  acknowledges funding from MIUR under the Grant No. PRIN 2017-MB8AEZ. S.B. acknowledges partial support from the project PGC2018-097585-B-C22. 
This work used the 2015 public version of the Munich model of galaxy formation and evolution: \LGalaxies. The source code and a full description of the model are available at http://galformod.mpa-garching.mpg.de/public/LGalaxies/. Finally, we thank the anonymous referee for the many suggestions that improved the quality of the paper.


\section*{DATA AVAILABILITY}

The simulated data underlying this article will be shared on reasonable request to the corresponding author.


\bibliographystyle{mnras}
\bibliography{references} 

\bsp	
\label{lastpage}
\end{document}